\documentclass[prd,showpacs,amsmath,amssymb,superscriptaddress,floatfix,nofootinbib,10pt]{revtex4}
\usepackage[utf8]{inputenc}
\usepackage[sort&compress]{natbib}
\usepackage{natbib}
\usepackage{ulem}
\usepackage{bm}
\usepackage{times}
\usepackage{amssymb,amsbsy,amsmath,amsfonts}
\usepackage{graphicx}
\usepackage{float}
\usepackage{color}
\usepackage{morefloats}
\usepackage{rotating}
\usepackage{srcltx}
\usepackage{slashed}
\usepackage{subfigure}
\usepackage{multirow}
\usepackage{verbatim}
\usepackage{hyperref}
\usepackage{tabularx}

\DeclareUnicodeCharacter{2212}{\textendash}
\begin{document}

\title{Strangeness $S = -2$ baryon-baryon interactions and femtoscopic correlation functions in covariant chiral effective field theory}

\author{Zhi-Wei Liu}
\affiliation{School of Physics, Beihang University, Beijing, 102206, China}

\author{Kai-Wen Li}
\affiliation{Medical Management Department, CAS Ion Medical Technology Co., Ltd., Beijing, 100190, China}
\affiliation{Beijing Advanced Innovation Center for Big Data-Based Precision Medicine, School of Medicine and Engineering, Beihang University, Key Laboratory of Big Data-Based Precision Medicine (Beihang University), Ministry of Industry and Information Technology, Beijing, 100191, China}
\affiliation{School of Physics, Beihang University, Beijing, 102206, China}

\author{Li-Sheng Geng}
\email[E-mail: ]{lisheng.geng@buaa.edu.cn}
\affiliation{School of Physics, Beihang University, Beijing, 102206, China}
\affiliation{Beijing Key Laboratory of Advanced Nuclear Materials and Physics, Beihang University, Beijing, 102206, China}
\affiliation{School of Physics and Microelectronics, Zhengzhou University, Zhengzhou, Henan, 450001, China}

\begin{abstract}

We study the baryon-baryon interactions with strangeness $S = -2$ and corresponding momentum correlation functions in leading order covariant chiral effective field theory. The relevant low energy constants are determined by fitting to the latest HAL QCD simulations, taking into account all the coupled channels. Extrapolating the so-obtained strong interactions to the physical point and considering both quantum statistical effects and the Coulomb interaction, we calculate the $\Lambda\Lambda$ and $\Xi^-p$ correlation functions with a spherical Gaussian source and compare them with the recent experimental data. We find  remarkable agreement between our predictions and the experimental measurements by using the source radius determined in proton-proton correlations, which demonstrates the consistency between theory, experiment, and lattice QCD simulations. Moreover, we predict the $\Sigma^+\Sigma^+$, $\Sigma^+\Lambda$, and $\Sigma^+\Sigma^-$ interactions and corresponding momentum correlation functions. We further investigate the influence of the source shape and size  of the hadron pair on the  correlation functions studied and show that  the current data are not very sensitive to the source shape. Future experimental measurement of the predicted momentum correlation functions will provide a non-trivial test of not only SU(3) flavor symmetry and its breaking but also the baryon-baryon interactions derived in covariant chiral effective field theory.

\end{abstract}


\maketitle

\section{Introduction}

Studies of hyperon-nucleon (YN) and hyperon-hyperon (YY) interactions play a significant role in enriching our knowledge about the residual strong interaction. In particular, the interactions in the strangeness $S = -2$ sector ($\Lambda\Lambda, \Xi N, \Sigma\Lambda, \Sigma\Sigma$) have gained more and more attention in many studies of current interest in nuclear physics and nuclear astrophysics, such as the H-dibaryon~\cite{Jaffe1977PRL38.195, NPLQCD2011PRL106.162001, HALQCD2011PRL106.162002, Shanahan2011PRL107.092004, Green2021PRL127.242003}, double $\Lambda$ and $\Xi$ hypernuclei~\cite{Takahashi2001PRL87.212502, Hiyama2020PRL124.092501, JPARC2021PRL126.062501}, and the so-called ``hyperon puzzle'' related to the maximum neutron star masses~ \cite{Lonardoni2014PRC89.014314, Maslov2015PLB748.369, Oertel2015JPG42.075202, Lim2015IJMPE24.1550100}. Different from the nucleon-nucleon case, where a large amount of high-quality scattering data exist, there are limited scattering data in the $S = -1$ sector due to the short life times of hyperons~\cite{Engelmann1966PL21.587, Alexander1968PR173.1452, Sechi1968PR75.1735, Hepp1968ZP214.71, Eisele1971PLB37.204, CLAS2021PRL127.272303}. In the $S = -2$ sector, direct scattering data are almost non-existent except for a few reaction cross-sections of relatively poor quality~\cite{Ahn2006PLB633.214}. Meanwhile, the descriptions of the YN and YY interactions by both phenomenological models and chiral effective field theories (ChEFT) also suffer from the lack of experimental constraints~\cite{Stoks1999PRC59.3009, Polinder2006NPA779.244, Fujiwara2007PPNP58.439, Haidenbauer2010PLB684.275, Haidenbauer2013NPA915.0375, Haidenbauer2016NPA954.273, KaiWen2018CPC42.014105, KaiWen2018PRC98.065203, Liu2021PRC103.025201}.

In the last few years, there has been remarkable progress in studies of the $S = -2$ baryon-baryon interactions both experimentally and theoretically. In particular, it was demonstrated that the momentum correlations between a pair of hadrons produced in heavy-ion collisions not only depend on quantum statistical effects and the space-time structure of the emitting source~\cite{Goldhaber1960PR120.300, Fung1978PRL41.1592, Zajc1984PRC29.2173, Bamberger1988PLB203.320, Podgoretsky1989FECAY20.628, Abbott1992PRL69.1030, Barrette1994PLB333.33, Wiedemann1999PR319.145, CMS2010PRL105.032001} but also are sensitive to the final-state interactions of the emitted hadron pair~\cite{Koonin1977PLB70.43, Gyulassy1979PRC20.2267, Lednicky1981YF35.1316, Pratt1990PRC42.2646, Bauer1992ARNPS42.77}. Because of the abundant hyperons produced in relativistic heavy-ion collisions and the excellent capabilities of detectors to identify particles and measure their momenta, the measurements of momentum correlation functions have become invaluable to reveal the precise dynamics of the strong interactions between a pair of hadrons, including meson-meson~\cite{ALICE2017PRC96.064613, ALICE2017PLB774.64, ALICE2019PLB790.22, ALICE2021arXiv2111.06611, ALICE2021PLB813.136030}, meson-baryon~\cite{ALICE2020PRL124.092301, ALICE2021PRC103.055201, ALICE2021PRL127.172301}, and (anti)baryon-(anti)baryon~\cite{STAR2015PRL114.022301, STAR2015Nature527.345, HADES2016PRC94.025201, ALICE2019PRC99.024001, ALICE2019PLB797.134822, STAR2019PLB790.490, ALICE2019PRL123.112002, ALICE2020PLB811.135849, ALICE2020PLB805.135419, ALICE2020PLB802.135223, ALICE2020Nature588.232, ALICE2021ARNPS71.377, ALICE2021arXiv2104.04427, ALICE2021arXiv2105.05190, STAR2022EPJWC295.11015}. The measurements of momentum correlation functions have also triggered a large amount of related theoretical studies~\cite{Ohnishi2000NPA670.297, Ohnishi2015PRC91.024916, Ohnishi2016PRC94.031901, Ohnishi2016NPA954.294, Ohnishi2017PPNP95.279, ALICE2018EPJC78.394, Ohnishi2020PRC101.015201, Ohnishi2020PRL124.132501, Ohnishi2021PRC105.014915, Ohnishi2021FBS62.42, Ohnishi2021PRC103.065205, Haidenbauer2019NPA981.1, Haidenbauer2020EPJA56.184, Haidenbauer2022PLB829.137074, Graczykowski2021PRC104.054909, Silverio2021arXiv2110.15455, Mrowczynski2021PRC104.024909}.

Meanwhile, with ever-growing computing power and evolving numerical algorithms, it has become possible to derive the YN and YY interactions from first principles lattice QCD simulations~\cite{NPLQCD2007NPA794.62, Nemura2018EPJWebConf175.05030, Ishii2018EPJWebConf175.05013, Doi2018EPJWebConf175.05009, Sasaki2020NPA998.121737, NPLQCD2021PRD103.054508}. In the $S = -2$ sector, the HAL QCD Collaboration has performed lattice QCD simulations for the $S$-wave $\Lambda\Lambda$ and $\Xi N$ interactions with an almost physical pion mass ($m_\pi = 146$ MeV)~\cite{Sasaki2020NPA998.121737}. Recently, a covariant ChEFT was proposed to describe the nucleon-nucleon interaction~\cite{Ren2018CPC42.014103, Xiao2019PRC99.024004, Xiao2020PRC102.054001, Bai2020PLB809.135745, Bai2021arXiv2105.06113, Wang2022PRC105.014003, Lu2022PRL128.142002, Ren2021CPL38.062101, Wang2021CPC45.054101}. As an extension of the theoretical framework to the $u, d, s(c)$ flavor space, the covariant ChEFT has also been applied in describing the YN and YY systems with strangeness ranging from $-1$ to $-4$ \cite{KaiWen2016PRD94.014029, KaiWen2018CPC42.014105, Song2018PRC97.065201, KaiWen2018PRC98.065203, Song2022PRC105.035203, Liu2021PRC103.025201}, and $\Lambda_c N$ system~\cite{Song2020PRC102.065208, Song2021arXiv2104.02380}. Given the latest experimental progress in Femtography, it is of critical importance to compare the covariant chiral YN and YY interactions constrained by the latest lattice QCD simulations with the measured correlation functions, especially in the $S = -2$ sector, which is the main purpose of the present work.

In addition, the $\Sigma\Sigma~(I = 2)$ interaction belongs to the same SU(3) irreducible representation ``27'' as the $NN~(I = 1)$, $\Sigma N~(I = 3/2)$, $\Xi N~(I = 3/2)$ and $\Xi\Xi~(I = 1)$ $^1S_0$ interactions and does not mix with other two-baryon channels. In principle, the behavior of these five channels is the same in the SU(3) symmetric limit, but in practice SU(3) flavor symmetry is broken due to the different masses of octet baryons and pseudoscalar mesons. Thus it offers an ideal platform to test SU(3) flavor symmetry and its breaking~\cite{Haidenbauer2015EPJA51.17, Liu2021PRC103.025201, NPLQCD2021PRD103.054508}. Moreover, there are still some open questions in the theoretical predictions for the $\Sigma\Sigma~(I = 2)$ interaction, e.g., whether the interaction is strong enough to generate a bound state as in the Nijmegen model~\cite{Stoks1999PRC59.3009}, or a virtual state as in the SU(6) quark cluster model (fss2)~\cite{Fujiwara2007PPNP58.439}, or no bound state as in ChEFTs~\cite{Haidenbauer2015EPJA51.17, KaiWen2018PRC98.065203}. Given this situation, it is important to obtain experimental information about the $\Sigma\Sigma~(I = 2)$ interaction, e.g., from the measurements of momentum correlation functions.

It is worthwhile to note that a recent study by Kamiya et al. used the aforementioned HAL QCD potential~\cite{Sasaki2020NPA998.121737} directly to derive the correlation functions of $\Lambda\Lambda$ and $\Xi^-p$ and ruled out the existence of a bound (quasibound) state in the $\Lambda\Lambda$ ($\Xi N$) channel~\cite{Ohnishi2021PRC105.014915}. Their analysis considered simultaneously the (incomplete) coupled-channel effects, the threshold difference, and the strong and Coulomb interaction. In the present work, although our main purpose is to test the covariant chiral $S=-2$ baryon-baryon interactions which are based on the HAL QCD phase shifts (not the potentials), we have made a number of improvements regarding the calculation of correlation functions: 1) the complete coupled-channel effects are taken into account, especially those of $\Sigma\Lambda$ and $\Sigma\Sigma$ channels that are neglected in Ref.~\cite{Ohnishi2021PRC105.014915}; 2) the strong interactions extrapolated to the physical point are used to evaluate the correlation function, while the two-body interactions used by Kamiya et al. are for $m_\pi = 146$ MeV;~\footnote{Replacing $m_\pi$ and $m_K$ in the parametrization of the HALQCD potential by the physical masses, Kamiya et al. found that due to the slightly increased attraction, the scattering length becomes slightly larger, although its central value is still located within the error bar of the lattice QCD one.}, 3) we predict the $\Sigma^+\Sigma^+$ correlation function, which can be used to test the SU(3) flavor symmetry and its breaking; 4) we predict the $\Sigma^+\Lambda$ and $\Sigma^+\Sigma^-$ correlation functions for the first time.

The paper is organized as follows. In Sec.~II we briefly explain how to evaluate two-hadron momentum correlation functions. In Sec.~III we first update the covariant chiral $S = -2$ baryon-baryon interactions by fitting to the latest HAL QCD simulations. Secondly, the $\Lambda\Lambda$ and $\Xi^-p$ correlation functions are analyzed in detail and compared with the experimental data. Next, the predictions for the $\Sigma\Sigma~(I = 2)$ phase shifts and $\Sigma^+\Sigma^+$, $\Sigma^+\Lambda$, and $\Sigma^+\Sigma^-$ correlation functions are presented. Finally, the influence of the source shape and size on the correlation functions is examined in detail. This article ends with a short summary and outlook.

\section{Theoretical Framework}\label{section:framework}
Two-hadron momentum correlation functions can be computed by the Koonin–Pratt (KP) formula \cite{Koonin1977PLB70.43, Pratt1990PRC42.2646, Bauer1992ARNPS42.77},
\begin{subequations}
  \begin{align}
    C(\boldsymbol{p}_1,\boldsymbol{p}_2)
    &=\frac{\int{\rm d}^4x_1{\rm d}^4x_2~S_1(x_1,\boldsymbol{p}_1)S_2(x_2,\boldsymbol{p}_2)~|\Psi^{(-)}(\boldsymbol{r},\boldsymbol{k})|^2}{\int{\rm d}^4x_1{\rm d}^4x_2~S_1(x_1,\boldsymbol{p}_1)S_2(x_2,\boldsymbol{p}_2)}\label{Eq:KP1}\\
    &\simeq\int{\rm d}\boldsymbol{r}~S_{12}(r)~|\Psi^{(-)}(\boldsymbol{r},\boldsymbol{k})|^2,\label{Eq:KP2}
  \end{align}
\end{subequations}
where $S_i(x_i,\boldsymbol{p}_i)~(i=1,2)$ is the single-particle source function of hadron $i$ with momentum $\boldsymbol{p}_i$. $\Psi^{(-)}$ denotes the relative wave function with the relative coordinate $\boldsymbol{r}$ and the relative momentum $\boldsymbol{k}=(m_2\boldsymbol{p}_1-m_1\boldsymbol{p}_2)/(m_1+m_2)$ in the center-of-mass (c.m.) frame, in which the effects of final-state interactions are embedded. If the time difference of the particle emission and the momentum dependence of the source function can be neglected, Eq.~\eqref{Eq:KP2} can be derived by integrating out the c.m. coordinates from Eq.~\eqref{Eq:KP1}, where $S_{12}(r)$ is the normalized source function of the pair. Following the standard practice~\cite{ALICE2019PLB797.134822, ALICE2019PRL123.112002, ALICE2020Nature588.232}, we assume a static and spherical Gaussian source with a single parameter $R$, namely $S_{12}(r) = {\rm exp}(-r^2/4R^2)/(2\sqrt{\pi}R)^3$.

In the present work, due to the dominant role of $S$-wave interactions in the low-momentum region, we assume that only the $S$-wave component of the relative wave function is modified by final-state interactions. For a non-identical two-particle system experiencing only strong interactions, the relative wave function in the two-body outgoing state can be written as \cite{Ohnishi2016NPA954.294}
\begin{align}\label{Eq:WF}
  \Psi^{(-)}_S(\boldsymbol{r},\boldsymbol{k})=e^{i\boldsymbol{k}\cdot\boldsymbol{r}}-j_0(kr)+\psi_0(r,k),
\end{align}
where the spherical Bessel function $j_0$ represents the $l = 0$ component of the non-interacting wave function, and $\psi_0$ denotes the $l = 0$ scattering wave function affected by the strong interaction. The scattering wave function $\psi_0$ is matched asymptotically to the boundary condition \cite{Ohnishi2016NPA954.294},
\begin{align}\label{Eq:asymptote1}
  \psi_0(r, k)\overset{r\to\infty}{\longmapsto}\frac{1}{2ikr}\left[e^{ikr}-e^{-2i\delta}e^{-ikr}\right],
\end{align}
where $\delta$ represent the phase shifts. Substituting the relative wave function \eqref{Eq:WF} into the KP formula, the correlation function then becomes 
\begin{align}\label{Eq:CF}
  C(k)\simeq1+\int_0^\infty4\pi r^2{\rm d}r~S_{12}(r)~\left[|\psi_0(r,k)|^2-|j_0(kr)|^2\right].
\end{align}
It should be emphasized that since the measured correlation functions are spin-averaged, the theoretical correlation functions should also be averaged over the total spin of the hadron pair with appropriate weights ($1/4$ for the spin-singlet and $3/4$ for the spin-triplet $S$-wave states), namely $C_s(k)/4+3C_t(k)/4$. 

In general, the scattering wave function can be obtained by solving the Schr\"odinger equation in coordinate space or the Lippmann-Schwinger (LS) / Kadyshevsky equation in momentum space~\cite{Ohnishi2015PRC91.024916, ALICE2018EPJC78.394, Haidenbauer2019NPA981.1}. In covariant ChEFT, the leading order (LO) four-baryon contact potentials are momentum-dependent (non-local) owing to the retention of the small component in the Dirac spinor \cite{Ren2018CPC42.014103, KaiWen2018CPC42.014105}. For our purpose it is convenient to first obtain the reaction amplitude $T$ by solving the LS / Kadyshevsky equation, and then derive the scattering wave function using the relation $|\psi\rangle=|\varphi\rangle+G_0T|\varphi\rangle$, where $G_0$ and $|\varphi\rangle$ represent the free propagator and the free wave function, respectively. In this work, we follow the formalism and conventions of Ref. ~\cite{Haidenbauer2019NPA981.1} and calculate the scattering wave function in the following way,
\begin{align}\label{Eq:Fourier_Bessel}
  \widetilde{\psi}_{\beta\alpha;l}(r)=\delta_{\beta\alpha}j_l(k_\alpha r)+\frac{1}{\pi}\int {\rm d}qq^2~\frac{T_{\beta\alpha;l}(q,k_\alpha;\sqrt{s})\cdot j_l(qr)}{\sqrt{s}-E_{\beta, 1}(q)-E_{\beta, 2}(q)+i\varepsilon},
\end{align}
where $T_{\beta\alpha;l}(q,k_\alpha;\sqrt{s})$ is the half-off-shell reaction amplitude, and the subscripts $\alpha$ and $\beta$ denote the incoming and outgoing channels, respectively. The total energy of the baryon-baryon system is defined as $\sqrt{s} = E_{\alpha, 1}(k_\alpha) + E_{\alpha, 2}(k_\alpha)$, where $E_{\gamma, i}(k)=\sqrt{m_{\gamma, i}^2 + k^2}~(\gamma = \alpha, \beta; i = 1, 2)$. The asymptotic scattering wave function has the following form
\begin{align}\label{Eq:asymptote2}
  \widetilde{\psi}_{\beta\alpha;l}(r)\overset{r\to\infty}{\longmapsto}\sqrt{\frac{\rho_\beta(k_\beta)}{\rho_\alpha(k_\alpha)}}\left[\delta_{\beta\alpha}j_l(k_\alpha r)-i\sqrt{\rho_\beta(k_\beta)\rho_\alpha(k_\alpha)}\cdot T_{\beta\alpha;l}(k_\beta, k_\alpha; \sqrt{s})\cdot h_l^{(1)}(k_\beta r)\right],
\end{align}
where $h_l^{(1)}$ represents the Hankel function of the first kind, and the phase-space factor is $\rho_\gamma(k) = k~E_{\gamma, 1}(k)~E_{\gamma, 2}(k)/(E_{\gamma, 1}(k) + E_{\gamma, 2}(k))$. The scattering wave function is abbreviated as $\widetilde{\psi}_l$ for the single-channel case ($\beta = \alpha$). Although there is a shift of an overall phase $\psi_l=e^{-2i\delta}~\widetilde{\psi}_l$, this difference does not change the modulus squared of the wave function in Eq.~\eqref{Eq:CF}. It should be noted that different from Refs.~\cite{Haidenbauer2019NPA981.1, Haidenbauer2020EPJA56.184, Haidenbauer2022PLB829.137074}, the reaction amplitude $T_{\beta\alpha;l}$ in this work is obtained by solving the coupled-channel Kadyshevsky equation~\cite{Kadyshevsky1968NPB6.125}, instead of the non-relativistic LS equation. The crucial distinction between the LS and Kadyshevsky equations is the propagator, which leads to the fact that the latter is less dependent on the momentum cutoff \cite{KaiWen2016PRD94.014029}. On the other hand, for calculating the wave functions, we have checked that the effects from different propagators can be neglected.

As mentioned in Ref.~\cite{Haidenbauer2019NPA981.1}, we can consider coupled-channel effects by replacing the modulus squared of the wave function in Eq.~\eqref{Eq:CF} with
\begin{align}\label{Eq:Coupling}
  |\psi_0(r,k)|^2\rightarrow\sum_\beta\omega_\beta|\widetilde{\psi}_{\beta\alpha;0}(r)|^2,
\end{align}
where the sum runs over all possible coupled channels, and $\omega_\beta$ is the weight for the individual components of the multi-channel wave function. In fact, the same source function has been assumed for the different channels to obtain the above formula.

\section{Results and Discussions}

\subsection{Fits to the state-of-the-art $S = -2$ lattice QCD data}\label{subsection:fitting}

In the covariant ChEFT~\cite{KaiWen2018PRC98.065203}, at LO there are 12 independent low-energy constants (LECs) for the $S$-wave contact potentials in the $S = -2$ system. So far, these LECs cannot be directly determined due to the lack of scattering data. In addition, one cannot assume SU(3) symmetry and simply relate the LECs to those of the $S = 0$ or $-1$ sector because of SU(3) symmetry breaking between sectors of different strangeness, as explicitly demonstrated in Refs.~\cite{Haidenbauer2015EPJA51.17, KaiWen2018CPC42.014105, Liu2021PRC103.025201}. Therefore we use the results of state-of-the-art lattice QCD simulations to fix these LECs.

Recently, the HAL QCD Collaboration reported lattice QCD simulations of the $S = -2$ baryon-baryon interactions near the physical point (with a pion mass $m_\pi = 146$ MeV)~\cite{Sasaki2020NPA998.121737}. They obtained the $\Lambda\Lambda$ and $\Xi N~(I = 0)~^1S_0$ phase shifts and the inelasticity~\footnote{It should be noted that they worked with the effective $\Lambda\Lambda-\Xi N$ coupled channels, instead of the full $\Lambda\Lambda-\Xi N-\Sigma\Sigma$ coupled channels.}, the $\Xi N~(I = 0)~^3S_1$ phase shifts, and the $\Xi N~(I = 1)~^1S_0$ and $^3S_1$ phase shifts~\footnote{They worked with the effective $\Xi N$ single channel, instead of the full $\Xi N-\Sigma\Lambda-\Sigma\Sigma$ coupled-channels.}. These results show better convergence than the preliminary results reported in 2018~ \cite{Sasaki2018EPJWebConf175.05010}, which have been used in our previous work~\cite{KaiWen2018PRC98.065203}. Hence we have to re-determine the 12 LECs by fitting to the latest lattice QCD data.

In the present work, we adopt the theoretical framework of Ref. ~\cite{KaiWen2018PRC98.065203}. For more details about the covariant ChEFT, especially for the YN and YY systems, we refer the reader to Refs.~\cite{KaiWen2018CPC42.014105, Song2018PRC97.065201, KaiWen2018PRC98.065203, Liu2021PRC103.025201, Song2022PRC105.035203}. To be consistent with Ref.~\cite{Liu2021PRC103.025201}, we consider cutoff values in the range of $550-700$ MeV. Since the HAL QCD method provides more reliable results with increasing imaginary-time distances $t/a$, bur for larger $t/a$ the statistical errors increase as well, we choose the intermediate $t/a = 12$ results to balance stability and reliability. We fit to the low-energy $\Lambda\Lambda~(I = 0)~^1S_0$ phase shifts with $E_{\rm c.m.}\leqslant10~{\rm MeV}$, the $\Lambda\Lambda$ and $\Xi N~(I = 0)~^1S_0$ phase shifts and the inelasticity with $30~{\rm MeV}\leqslant E_{\rm c.m.}\leqslant40~{\rm MeV}$, the $\Xi N~(I = 0)~^3S_1$, $\Xi N~(I = 1)~^1S_0$ and $^3S_1$ phase shifts with $E_{\rm c.m.}\leqslant10~{\rm MeV}$. Different from Refs.~\cite{Aoki2011PJASB87.509, Aoki2013PRD87.034512, Sasaki2020NPA998.121737}, we consider all the coupled channels. The updated values of the LECs for different cutoffs $\Lambda_F$ are listed in Table~\ref{Tab:LECs_m2}, with a total $\chi^2/{\rm d.o.f. \approx 0.34}$.

\begin{table}[h]
  \centering
  \caption{Low-energy constants (in units of $10^4~{\rm GeV}^{-2}$) for cutoffs $\Lambda_F$ = 550 – 700 MeV in covariant ChEFT. These LECs are determined by fitting to the $\Lambda\Lambda$ and $\Xi N$ phase shifts provided by the HAL QCD Collaboration ($t/a = 12$) \cite{Sasaki2020NPA998.121737}.}
  \label{Tab:LECs_m2}
  \setlength{\tabcolsep}{5.2pt}
  \begin{tabular}{cccccccccrcccc}
    \hline
    \hline
    $\Lambda_F$&&$C_{1S0}^{\Lambda\Lambda}$ &$C_{1S0}^{\Sigma\Sigma}$ &$C_{3S1}^{\Lambda\Lambda}$ &$C_{3S1}^{\Sigma\Sigma}$ &$C_{3S1}^{\Lambda\Sigma}$ &$C_{1S0}^{4\Lambda}$ &$\hat{C}_{1S0}^{\Lambda\Lambda}$ &$\hat{C}_{1S0}^{\Sigma\Sigma}$ &$\hat{C}_{3S1}^{\Lambda\Lambda}$~~ &$\hat{C}_{3S1}^{\Sigma\Sigma}$ &$\hat{C}_{3S1}^{\Lambda\Sigma}$ &$\hat{C}_{1S0}^{4\Lambda}$\\
    \hline
    $550$  &&$  -0.0274$ &$-0.0412$ &$-0.0078$ &$0.0255$ &$0.0024$ &$  -0.0242$ &$2.3493$ &$2.5353$ &$1.3695$ &$1.0552$ &$  -0.0423$ &$1.9485$ \\
    $600$  &&$  -0.0175$ &$-0.0300$ &$-0.0076$ &$0.0472$ &$0.0026$ &$  -0.0176$ &$2.0832$ &$2.2246$ &$1.0521$ &$1.1759$ &$~~~0.0793$ &$1.8207$ \\
    $650$  &&$  -0.0049$ &$-0.0169$ &$-0.0070$ &$0.0720$ &$0.0026$ &$  -0.0075$ &$1.9847$ &$2.0755$ &$0.8493$ &$1.1768$ &$~~~0.0793$ &$1.8207$ \\
    $700$  &&$~~~0.0089$ &$-0.0053$ &$-0.0064$ &$0.1049$ &$0.0026$ &$~~~0.0066$ &$1.8566$ &$1.8869$ &$0.7072$ &$1.1768$ &$~~~0.0793$ &$1.8206$ \\
    \hline
    \hline
  \end{tabular}
\end{table}

\begin{figure}[h]
  \centering
  \includegraphics[width=0.95\textwidth]{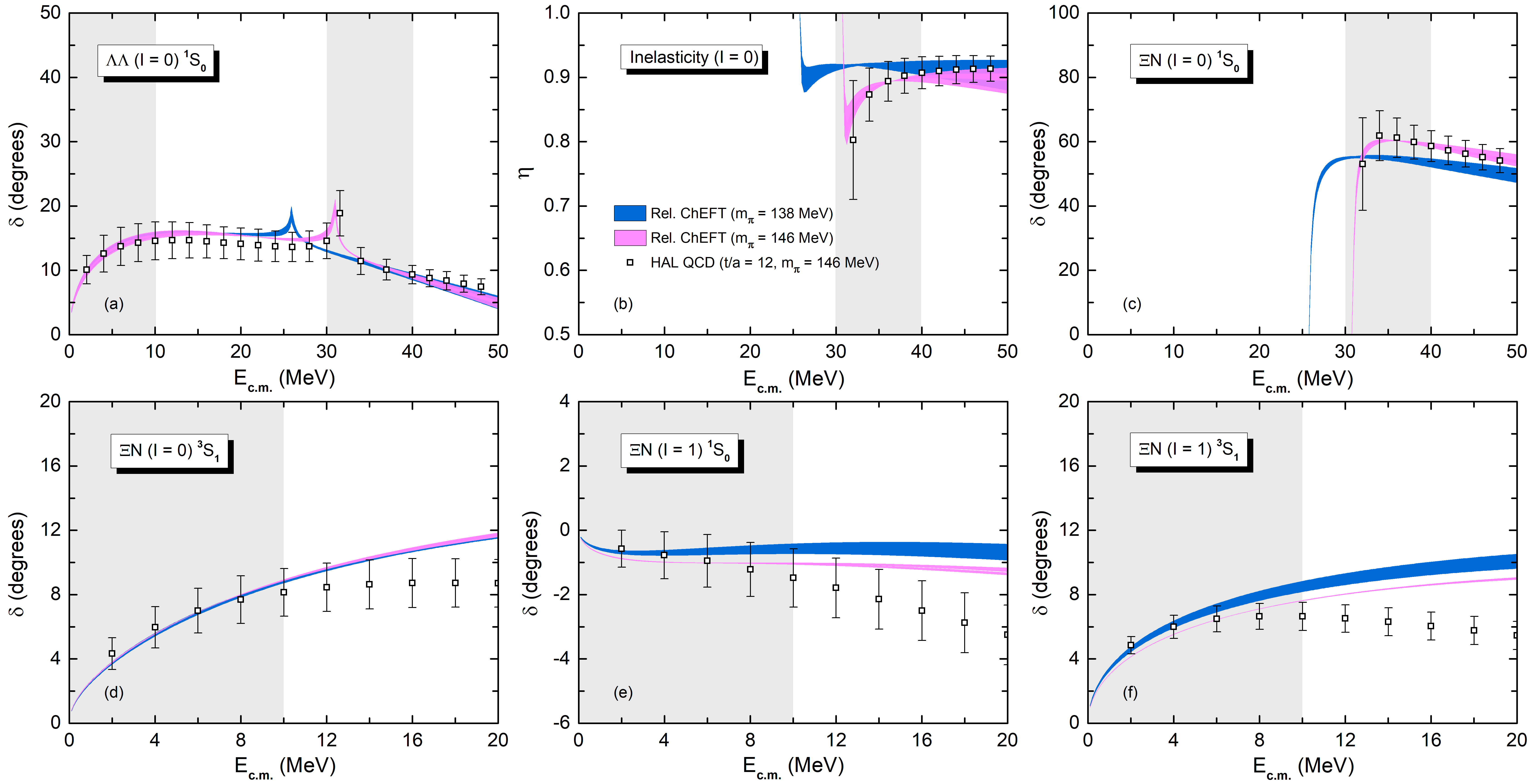}
  \caption{(color online) (a-c) $\Lambda\Lambda$ and $\Xi N$ ($I = 0$) $^1S_0$ phase shifts and the inelasticity, (d) $\Xi N$ ($I = 0$) $^3S_1$ phase shift, (e-f) $\Xi N$ ($I = 1$) $^1S_0$ and $^3S_1$ phase shifts as functions of the c.m. kinetic energy $E_{\rm c.m.}$. The LECs are fitted to the HAL QCD results ($t/a = 12$) in the gray regions, taken from Ref. \cite{Sasaki2020NPA998.121737}. The uncertainties due to the variation of the cutoff in the range of $\Lambda_F$ = 550 – 700 MeV are shown by the shaded bands.
  }\label{Fig:PS_fitting}
\end{figure}

The fitted and extrapolated $S$-wave phase shifts and inelasticity are shown in Fig.~\ref{Fig:PS_fitting}. The fitted results are shown as light magenta bands and the extrapolated results to the physical point are shown as dark blue ones. The bands reflect the variation of the cutoff in the range of $550-700$ MeV. As mentioned above, only the low-energy lattice QCD data in the shadowed regions are fitted. It is clear that the $\Lambda\Lambda$ and $\Xi N~(I = 0)~^1S_0$ phase shifts and the inelasticity are in very good agreement with the lattice QCD data in the whole energy region (see panels a-c); the $\Xi N~(I = 0)~^3S_1$, $\Xi N~(I = 1)~^1S_0$ and $^3S_1$ phase shifts are also in good agreement with the lattice QCD data in the low-energy region, while the predicted results are slightly higher than the lattice QCD data in the high-energy region (see panels d-f), which can be improved at higher chiral orders. Here, we would like to remind the reader that the c.m. kinetic energy $E_{\rm c.m.} = 10$ MeV corresponds to the relative momentum $k \approx 105$ MeV for the $\Xi N$ system, and the inadequacy of the ChEFT in describing the lattice QCD data in the higher energy region does not affect the correlation function in any significant way in the region of our interest, i.e., $k < 300$ MeV. It is worth mentioning that since the lattice QCD simulations were obtained with almost physical pion masses, the extrapolated results to the physical point in panels (d-f) are close to the fitted phase shifts. However, there are visible differences between the fitted and extrapolated results due to the shift of the $\Xi N$ threshold [see panels (a-c)]. 

\subsection{$\Lambda\Lambda$ correlation function}\label{subsection:LambdaLambda}

First, we analyse the $\Lambda\Lambda$ correlation function computed with the $S = -2$ baryon-baryon interactions obtained above. For systems of two identical-particles, quantum statistical effects have to be taken into account, which requires symmetrizing or antisymmetrizing the relative wave function with respect to the exchange of the coordinates of two particles. For the $\Lambda\Lambda$ system with the strong interaction, the relative wave function can be decomposed into the symmetric component with even parity (spin-singlet) and the antisymmetric component with odd parity (spin-triplet), namely,
\begin{subequations}
  \begin{align}
    \Psi_{S,E}^{(-)}(\boldsymbol{r},\boldsymbol{k})&=\frac{\Psi_S^{(-)}(\boldsymbol{r},\boldsymbol{k})+\Psi_S^{(-)}(-\boldsymbol{r},\boldsymbol{k})}{\sqrt{2}}=\sqrt{2}\left[\cos(\boldsymbol{k}\cdot\boldsymbol{r})-j_0(kr)+\psi_0(r,k)\right],\label{Eq:WF_symmetry1}\\
    \Psi_{S,O}^{(-)}(\boldsymbol{r},\boldsymbol{k})&=\frac{\Psi_S^{(-)}(\boldsymbol{r},\boldsymbol{k})-\Psi_S^{(-)}(-\boldsymbol{r},\boldsymbol{k})}{\sqrt{2}}=\sqrt{2}i\sin(\boldsymbol{k}\cdot\boldsymbol{r}).\label{Eq:WF_symmetry2}
  \end{align}
\end{subequations}
Note that the modification due to the final-state interaction is forbidden for the antisymmetric wave function by the Pauli principle. Substituting the relative wave functions \eqref{Eq:WF_symmetry1} and \eqref{Eq:WF_symmetry2} into the KP formula with the appropriate spin weights, we obtain the following $\Lambda\Lambda$ correlation function
\begin{align}\label{Eq:CF_symmetry}
  C_{\Lambda\Lambda}(k)\simeq1-\frac{1}{2}e^{-4k^2R^2}+\frac{1}{2}\int_0^\infty4\pi r^2{\rm d}r~S_{12}(r)~\left[|\psi_0(r,k)|^2-|j_0(kr)|^2\right],
\end{align}
where the second term on the right-hand side represents the so-called quantum statistical effect, which suppresses the $\Lambda\Lambda$ correlation over the whole $k$ range. As explained in Sec. \ref{section:framework}, to consider coupled-channel effects, Eq.~\eqref{Eq:Coupling} should be substituted into the above correlation function. We note that there are already some studies on the $\Lambda\Lambda$ correlation function, but not all the coupled channels are taken into account in these works. For instance, the contributions of the $\Sigma^0\Sigma^0$ and $\Sigma^+\Sigma^-$ components are not considered in Refs~\cite{Haidenbauer2019NPA981.1, Ohnishi2021PRC105.014915}. In our study, all the coupled channels of the $\Lambda\Lambda$ system including $\Xi^0n$, $\Xi^-p$, $\Sigma^0\Sigma^0$, and $\Sigma^+\Sigma^-$ are considered. For the convenience of discussion, we assume the same source radius $R = 1.2$ fm and weight $\omega_\beta = 1$ for all the channels in this subsection, and also in Sec.~\ref{subsection:Xip}, \ref{subsection:SigmaSigma}.

\begin{figure}[h]
  \centering
  \includegraphics[width=0.44\textwidth]{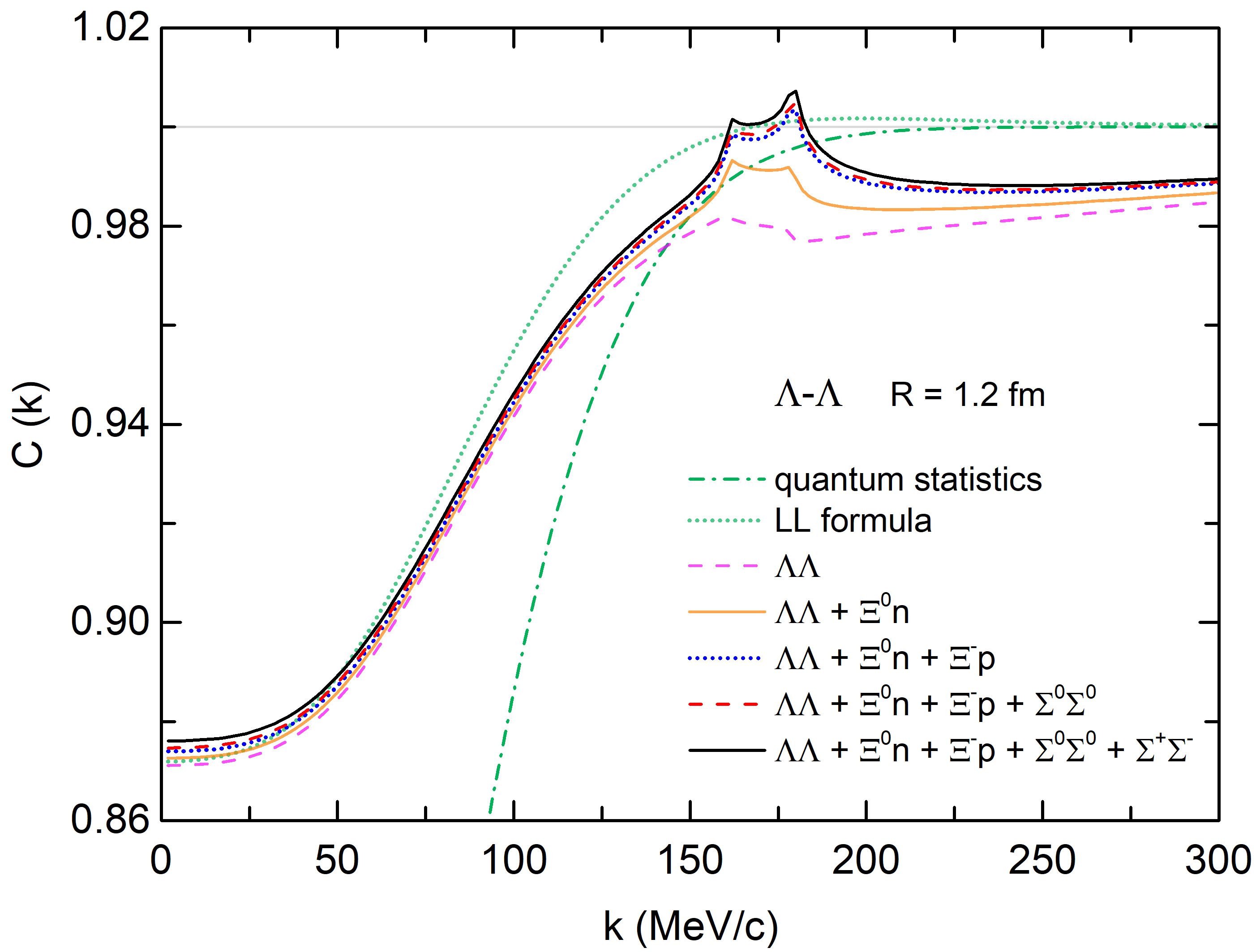}
  \caption{(color online) $\Lambda\Lambda$ correlation function as a function of the relative momentum $k$. The results are calculated with the covariant chiral baryon-baryon interactions (obtained with $\Lambda_F = 600$ MeV). The light magenta dashed line denotes the correlation function for which only the $\Lambda\Lambda$ wave function is taken into account, while the light orange solid line, the blue dotted line, the red dashed line and the black solid line denote the results in which the $(\Xi^0n, \Xi^-p, \Sigma^0\Sigma^0, \Sigma^+\Sigma^-)$ contributions are considered one by one, respectively. The results obtained by considering only quantum statistics and the Lednicky–Lyuboshitz model (see Refs. \cite{Lednicky1981YF35.1316, Ohnishi2016NPA954.294, Haidenbauer2019NPA981.1} for more details) are also shown for comparison. 
  }\label{Fig:CF_LambdaLambda}
\end{figure}

In Fig.~\ref{Fig:CF_LambdaLambda}, we show the full coupled-channel $C_{\Lambda\Lambda}$ (the black solid line), the contributions from different inelastic channels, the prediction of the Lednicky–Lyuboshitz (LL) model (the light green dotted line), as well as the case of pure quantum statistical effects (the green dash-dotted line). Obviously, due to the attractive strong interaction, there is a significant enhancement in $C_{\Lambda\Lambda}$ compared to the case of pure quantum statistical effects below $150$ MeV$/c$. It is interesting to see that the openings of the inelastic $\Xi^0 n$ and $\Xi^-p$ channels cause the appearance of two cusp-like structures at the corresponding thresholds, consistent with Refs.~\cite{Haidenbauer2019NPA981.1, Ohnishi2021PRC105.014915}. However, it would be a challenge to observe these cusp-like structures in future experiments due to the weak $\Lambda\Lambda-\Xi N$ coupling. The suppressed contributions from the $\Sigma^0\Sigma^0$ and $\Sigma^+\Sigma^-$ components are in line with expectations because the corresponding wave functions drop fast in the momentum region below the $\Sigma^0\Sigma^0$ and $\Sigma^+\Sigma^-$ thresholds, which justifies the neglect of the $\Sigma\Sigma$ contribution in Refs.~\cite{Haidenbauer2019NPA981.1, Ohnishi2021PRC105.014915}.. In addition, the full coupled-channel result can be approximated very well at the low-momentum region by the LL model, in which the scattering length $a_s^{\Lambda\Lambda} \approx -0.77$ fm and the effective range $r_s^{\Lambda\Lambda} \approx 3.83$ fm.

\subsection{$\Xi^-p$ correlation function}\label{subsection:Xip}

For systems of two charged particles, the contribution from the Coulomb interaction has to be taken into account, which is expected to play a significant role in the low-momentum region. For the $\Xi^-p$ system, considering the strong and Coulomb interactions, the relative wave function reads
\begin{align}\label{Eq:WF_Coulomb}
  \Psi_{SC}^{(-)}(\boldsymbol{r},\boldsymbol{k})=\phi^C(\boldsymbol{r},\boldsymbol{k})-\phi_0^C(kr)+\psi_0^{SC}(r,k),
\end{align}
where $\phi^C$ and $\phi_0^C$ denote the full Coulomb wave function and its $S$-wave component, respectively, and $\psi_0^{SC}$ denotes the scattering wave function including both the strong and Coulomb interactions, which can be obtained analogously to Eq.~\eqref{Eq:Fourier_Bessel}. In the present work, we adopt the Vincent-Phatak method to treat the Coulomb interaction in momentum space~\cite{Vincent1974PRC10.391, Holzenkamp1989NPA500.485,Haidenbauer2020EPJA56.184}. Substituting the relative wave function~\eqref{Eq:WF_Coulomb} into the KP formula, the $\Xi^-p$ correlation function is
\begin{align}\label{Eq:CF_Coulomb}
  C_{\Xi^- p}(k)\simeq\int{\rm d}\boldsymbol{r}~S_{12}(r)~|\phi^C(\boldsymbol{r},\boldsymbol{k})|^2+\int_0^\infty4\pi r^2{\rm d}r~S_{12}(r)~\left[|\psi_0^{SC}(r,k)|^2-|\phi_0^C(kr)|^2\right].
\end{align}
The final correlation function needs to be averaged over the spin of the hadron pair with the appropriate weights. Eq.~\eqref{Eq:Coupling} is also used to evaluate the coupled-channel effects. It is worth noting that $\Xi^-p$ in the spin-singlet channel can couple to $\Lambda\Lambda$, $\Xi^0n$, $\Sigma^0\Lambda$, $\Sigma^0\Sigma^0$, and $\Sigma^+\Sigma^-$, while in the spin-triplet channel it can couple to $\Xi^0n$, $\Sigma^0\Lambda$, and $\Sigma^+\Sigma^-$. It should be noted that in Refs~\cite{Haidenbauer2019NPA981.1, Ohnishi2021PRC105.014915} the contribution of the $\Sigma^0\Lambda$, $\Sigma^0\Sigma^0$ and $\Sigma^+\Sigma^-$ channels are neglected , while all the coupled channels of $\Xi^-p$ are taken into account in this study.

\begin{figure}[h]
  \centering
  \includegraphics[width=0.44\textwidth]{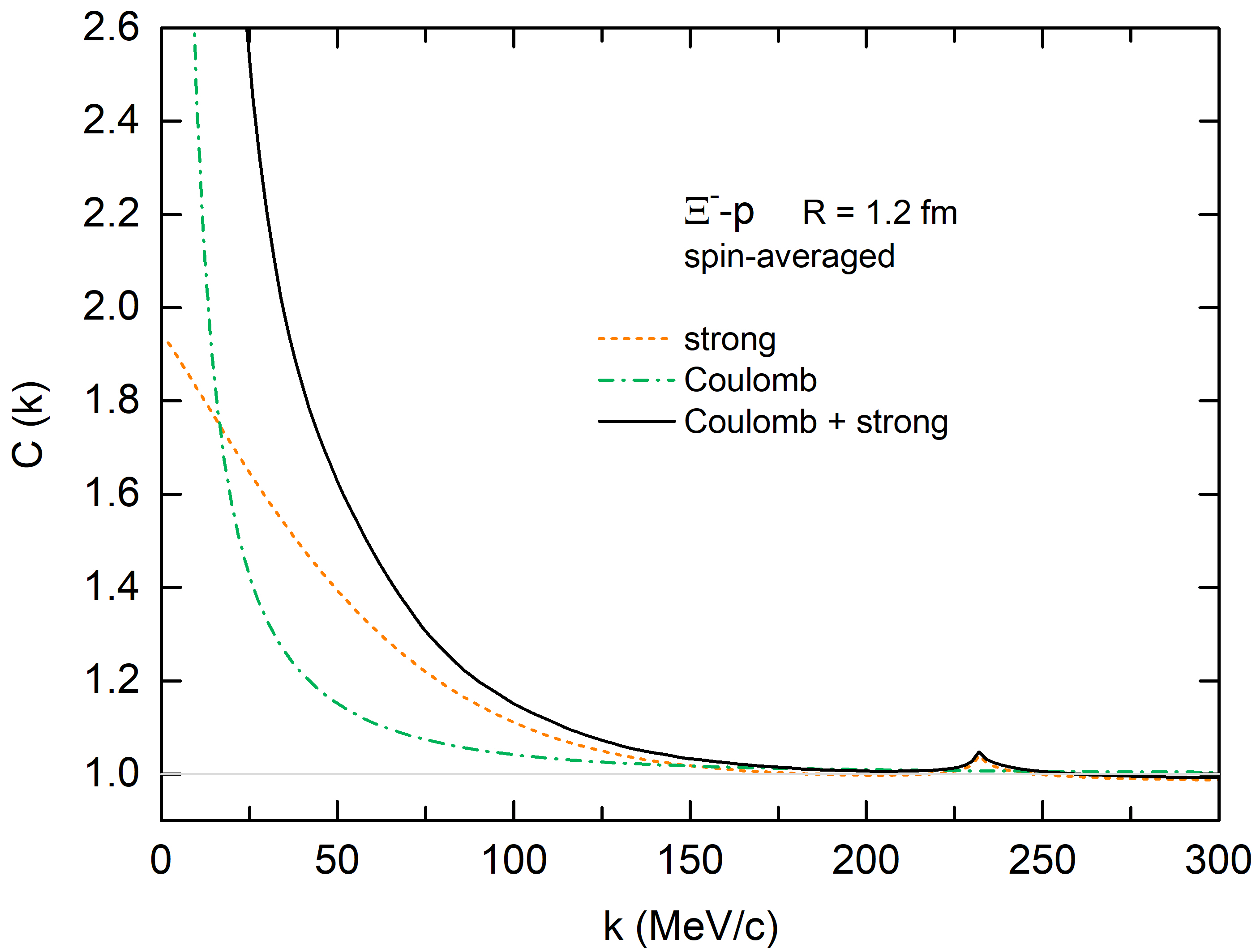}
  \caption{(color online) $\Xi^-p$ correlation function as a function of the relative momentum $k$. The results are calculated with the covariant chiral baryon-baryon interactions (obtained with $\Lambda_F = 600$ MeV). The solid (dash-dotted) line denotes the result with (without) the contribution of the strong interaction, while the short dashed line denotes the correlation function obtained considering only the strong interaction.
  }\label{Fig:CF_Ximp}
\end{figure}

\begin{figure}[h]
  \centering
  \includegraphics[width=0.95\textwidth]{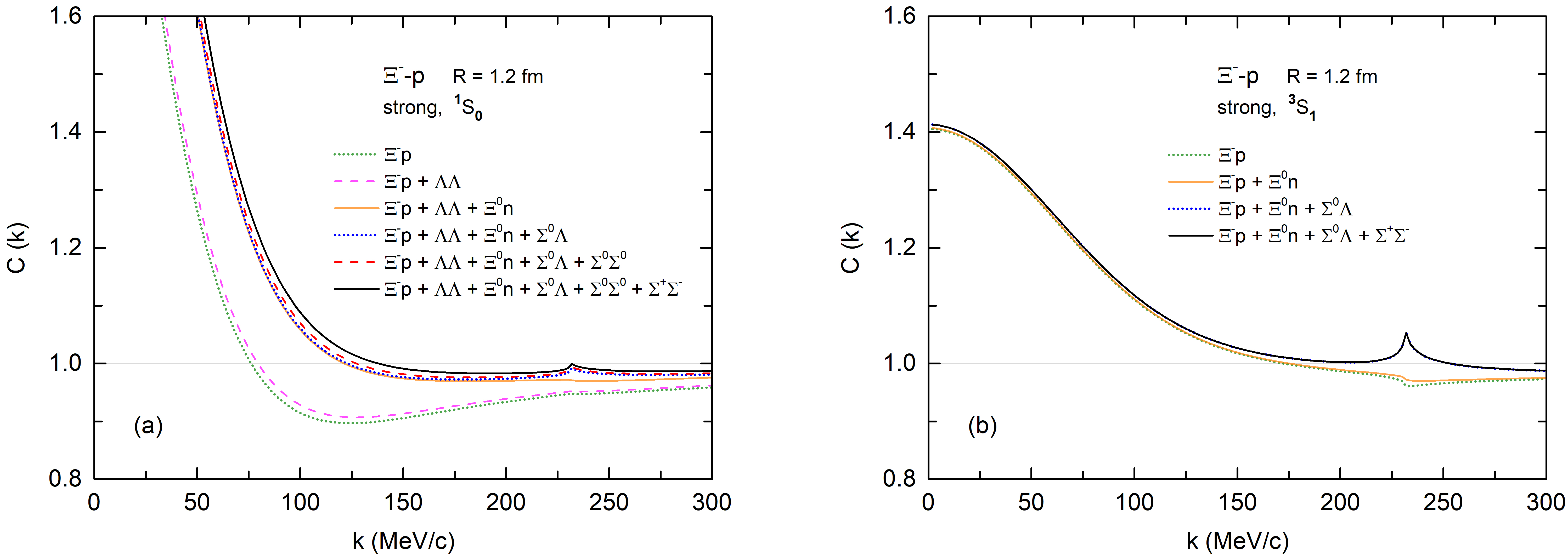}
  \caption{(color online) Breakdown of the strong interaction component of the $\Xi^-p$ correlation function, where $k$ is the relative momentum. (a) Spin-singlet part. The  green dotted line denotes the correlation function when only the $\Xi^-p$ $^1S_0$ wave function is taken into account, while the light magenta dashed line, the light orange solid line, the blue dotted line, the red dashed line and the black solid line denote the results in which the $(\Lambda\Lambda, \Xi^0n, \Sigma^0\Lambda, \Sigma^0\Sigma^0, \Sigma^+\Sigma^-)$ contributions are accumulated in order, respectively. (b) Spin-triplet part. The green dotted line denotes the correlation function for which only the $\Xi^-p$ $^3S_1$ wave function is taken into account, while the light orange solid line, the blue dotted line and the black solid line denote the results in which the $(\Xi^0n, \Sigma^0\Lambda, \Sigma^+\Sigma^-)$ contributions are added in order, respectively.
  }\label{Fig:CF_Ximp_singlet_triplet}
\end{figure}

In Fig.~\ref{Fig:CF_Ximp}, we show the full coupled-channel $C_{\Xi^-p}$ with and without the Coulomb interaction (the solid line and the short dashed line, respectively), as well as the case considering only the Coulomb attraction (the short dashed line). Compared with the correlation function obtained with the Coulomb attraction, we see a significant enhancement of $C_{\Xi^-p}$ below $150$ MeV$/c$, which is consistent with the strong interaction contribution in the low-momentum region. Note that an appreciable cusp-like structure shows up around $k \approx 230$ MeV$/c$, which corresponds to the opening of the $\Sigma^0\Lambda$ channel. To clarify the origin of this cusp-like structure and the influence of coupled-channel effects, we decompose the strong interaction contribution into the contributions from the $^1S_0$ (spin-singlet) and $^3S_1$ (spin-triplet) channels, and further decompose them according to the different inelastic channels, as shown in Fig.~\ref{Fig:CF_Ximp_singlet_triplet} (a) and (b). In accordance with the large negative scattering length in the $\Xi^-p$ $^1S_0$ channel, the correlation from the spin-singlet channel is also stronger. In addition, there is a more visible contribution in the spin-singlet channel from the $\Xi^-p-\Xi^0n$ coupled-channel, in agreement with Refs.~\cite{Haidenbauer2019NPA981.1, Ohnishi2021PRC105.014915}. It is clear that the cusp-like structure comes from the contribution of the $\Xi^-p-\Sigma^0\Lambda$ coupled channel, especially in the spin-triplet channel, which can be traced back to the stronger $\Xi^-p-\Sigma^0\Lambda$ coupling in the $^3S_1$ channel. The detailed structure might be observed in future high precision experiments, similar to the exploration of the $N\Lambda-N\Sigma$ system~\cite{ALICE2021arXiv2104.04427}.

\subsection{Comparison with experimental correlation functions}\label{subsection:experiment}

In order to test the covariant chiral interactions, we compare the theoretical $\Lambda\Lambda$ and $\Xi^-p$ correlation functions with the recent experimental data taken from $p$–Pb collisions at $\sqrt{s} = 5.02$ TeV \cite{ALICE2019PLB797.134822, ALICE2019PRL123.112002} and $p$–$p$ collisions at $\sqrt{s} = 13$ TeV \cite{ALICE2019PLB797.134822, ALICE2020Nature588.232}. For the sake of simplicity, we assume that the $pp$, $\Lambda\Lambda$, and $\Xi^-p$ pairs originate from a common source with a constant size at the same collision energy, and the value of the radius is determined via an independent analysis of proton-proton correlations, i.e., $R = 1.427$ fm for $p$–Pb collisions at $\sqrt{s} = 5.02$ TeV and $R = 1.182$ fm for $p$–$p$ collisions at $\sqrt{s} = 13$ TeV~\cite{ALICE2019PLB797.134822}. Although we have neglected the transverse mass dependence of the source size, the following qualitative results and conclusions are not expected to change. In addition, the more reasonable ratios of the weights $\omega_\beta/\omega_\alpha$ are estimated by applying the statistic model~\cite{Bazavov2014PRD90.094503, Borsanyi2014PLB730.99}, which read,
\begin{align}\label{Eq:Coupling_ratio}
  \frac{\omega_\beta}{\omega_\alpha}=\frac{\chi^\beta~{\rm exp}[(-m_{\beta, 1}-m_{\beta, 2})/T^*]}{\chi^\alpha~{\rm exp}[(-m_{\alpha, 1}-m_{\alpha, 2})/T^*]},
\end{align}
where the hadronization temperature $T^* = 154$ MeV. For the $\Lambda\Lambda$ system, the factor $\chi^{\Lambda\Lambda} = 1/2$; for the $\Xi N$ system, the factors $\chi^{\Xi N}_{J = 0} = 1/4$ and $\chi^{\Xi N}_{J = 1} = 3/4$~\cite{Ohnishi2021PRC105.014915}. The weight of the observed channel $\omega_\alpha$ is set at 1 because of the normalization of the source function~\cite{Ohnishi2020PRL124.132501}. The same statistic model is used to estimate the ratios of  $\omega_\beta/\omega_\alpha$ in Sec.~\ref{subsection:SigmapLambda_SigmapSigmam} and \ref{subsection:Source_shape_size_dependence. It should be noted that, due to the contamination from particle mis-identification, the feed-down effect, and the non-femtoscopic effect such as the minijet contribution in the experiments, we have to consider the following correction,}
\begin{align}\label{Eq:CF_fit}
  C_{\rm fit}(k)=(a~+~bk)~[1~+~\lambda(C_{\rm th}(k)~-~1)].
\end{align}
The theoretical correlation function $C_{\rm th}$ can be calculated in a similar way as shown in  Sec.~\ref{subsection:LambdaLambda} and \ref{subsection:Xip}. The pair purity probability $\lambda$ is generally used to estimate the influence from particle mis-identification and feed-down effects. As suggested by the experiments~\cite{ALICE2019PLB797.134822, ALICE2019PRL123.112002, ALICE2020Nature588.232}, $\lambda$ equals $0.239 (0.338)$ and $0.513 (1)$ for the $\Lambda\Lambda$ and $\Xi^-p$ pairs in $p$–Pb collisions at $\sqrt{s} = 5.02$ TeV ($p$–$p$ collisions at $\sqrt{s} = 13$ TeV), respectively. Here, we assume that other correlations feeding into the  channels of interests are flat. The parameters $a$ and $b$, which account for the non-femtoscopic effect, have to be determined by fitting to the experimental data. As we show later, $a$ is close to 1 and $b$ is very small. Therefore, we ignore the non-femtoscopic effect in this subsection. It should be stressed, as we have fixed all the relevant parameters, the theoretical results shown below are true predictions. The source shape and size dependence, as well as the non-femtoscopic effect are discussed in detail in Sec.~\ref{subsection:Source_shape_size_dependence}.

\begin{figure}[h]
  \centering
  \includegraphics[width=0.95\textwidth]{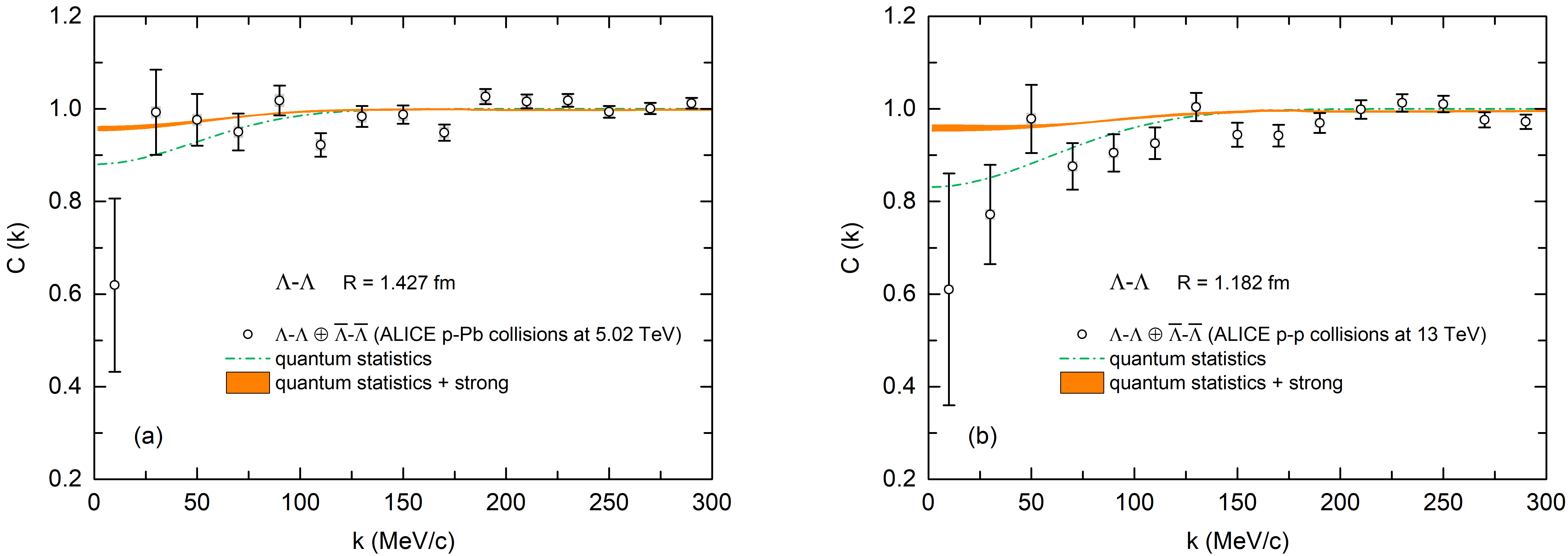}
  \caption{(color online) Theoretical $\Lambda\Lambda$ correlation function as a function of the relative momentum $k$, in comparison with the experimental data taken from $p$–Pb collisions at $\sqrt{s} = 5.02$ TeV \cite{ALICE2019PLB797.134822} and $p$–$p$ collisions at $\sqrt{s} = 13$ TeV \cite{ALICE2019PLB797.134822}. The theoretical results are calculated with the covariant chiral baryon-baryon interactions. The orange (shaded) bands reflect the variation of the cutoff in the range of $\Lambda_F$ = 550 – 700 MeV. The results obtained with pure quantum statistical effects  are also shown as green dash-dotted lines.
  }\label{Fig:CF_LambdaLambda_Exp}
\end{figure}

Fig.~\ref{Fig:CF_LambdaLambda_Exp} shows the theoretical $\Lambda\Lambda$ correlation function in comparison with the experimental data taken from $p$–Pb collisions at $\sqrt{s} = 5.02$ TeV~\cite{ALICE2019PLB797.134822} and $p$–$p$ collisions at $\sqrt{s} = 13$ TeV~\cite{ALICE2019PLB797.134822}. The theoretical results are calculated with the covariant chiral baryon-baryon interactions. The orange (shaded) bands reflect the variation of the cutoff in the range $\Lambda_F$ = 550 – 700 MeV. The contributions of pure quantum statistical effects are displayed as green dash-dotted lines. For the case of $p$–Pb collisions at $\sqrt{s} = 5.02$ TeV, the agreement with the experimental data indicates a weak $\Lambda\Lambda$ attraction, which rules out the existence of a deeply bound state. On the right panel, the theoretical result is also qualitatively similar to the experimental data but is larger in the low-momentum region.

\begin{figure}[h]
  \centering
  \includegraphics[width=0.95\textwidth]{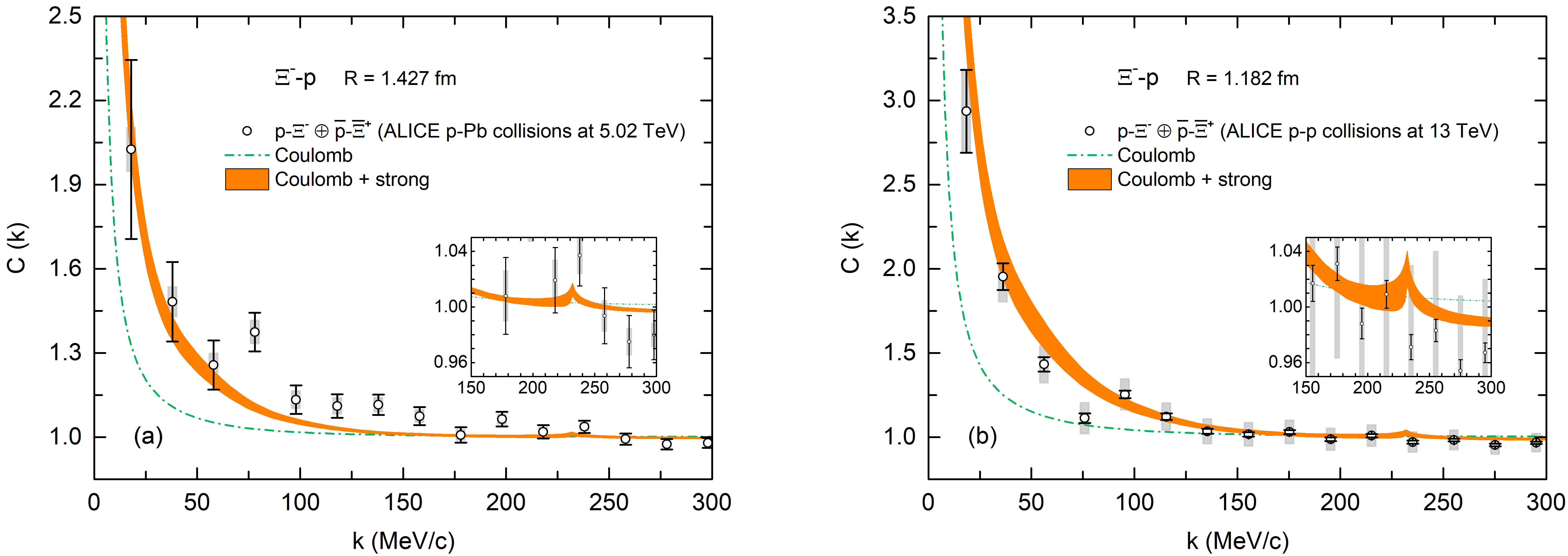}
  \caption{(color online) Theoretical $\Xi^-p$ correlation function as a function of the relative momentum $k$, in comparison with the experimental data taken from $p$–Pb collisions at $\sqrt{s} = 5.02$ TeV \cite{ALICE2019PRL123.112002} and $p$–$p$ collisions at $\sqrt{s} = 13$ TeV \cite{ALICE2020Nature588.232}. The theoretical results are calculated with the covariant chiral baryon-baryon interactions. The orange (shaded) bands reflect the variation of the cutoff in the range of $\Lambda_F$ = 550 – 700 MeV. The correlation functions obtained with only the Coulomb contribution are also shown as green dash-dotted lines.
  }\label{Fig:CF_Ximp_Exp}
\end{figure}

Similar comparisons are also performed for the $\Xi^-p$ correlation, as shown in Fig.~\ref{Fig:CF_Ximp_Exp}. The correlation functions with only the Coulomb contribution are also shown as green dash-dotted lines. At both collision energies, the enhancements over the cases of only the Coulomb contribution indicate the attractive nature of the $\Xi N$ strong interaction. As no free parameter is introduced, the remarkable agreements between our results and the experimental data demonstrate again the reliability of the covariant chiral potentials. In particular, the cusp-like structure around $k \approx 230$ MeV$/c$ may provide an opportunity for the direct experimental observation of the $\Xi N-\Sigma\Lambda$ coupled-channel effect in the $\Xi N$ system. 

It is interesting to note that the $\Lambda\Lambda$ and $\Xi^-p$ correlation functions obtained in the present work are qualitatively similar to those of Ref.~\cite{Ohnishi2021PRC105.014915} with two main differences. First, a cusp-like structure  appears around $k \approx 230$ MeV$/c$ in our $\Xi^-p$ correlation function, which is absent in Ref.~\cite{Ohnishi2021PRC105.014915} because they neglected the $\Sigma^0\Lambda$ channel. Second, we used the source size suggested by experiments while in Ref.~\cite{Ohnishi2021PRC105.014915} the source size is taken to be a free parameter and determined by fitting to the data.

\subsection{Predictions for the $\Sigma\Sigma~(I = 2)$ phase shifts and $\Sigma^+\Sigma^+$ correlation function}\label{subsection:SigmaSigma}

\begin{figure}[h]
  \centering
  \includegraphics[width=0.44\textwidth]{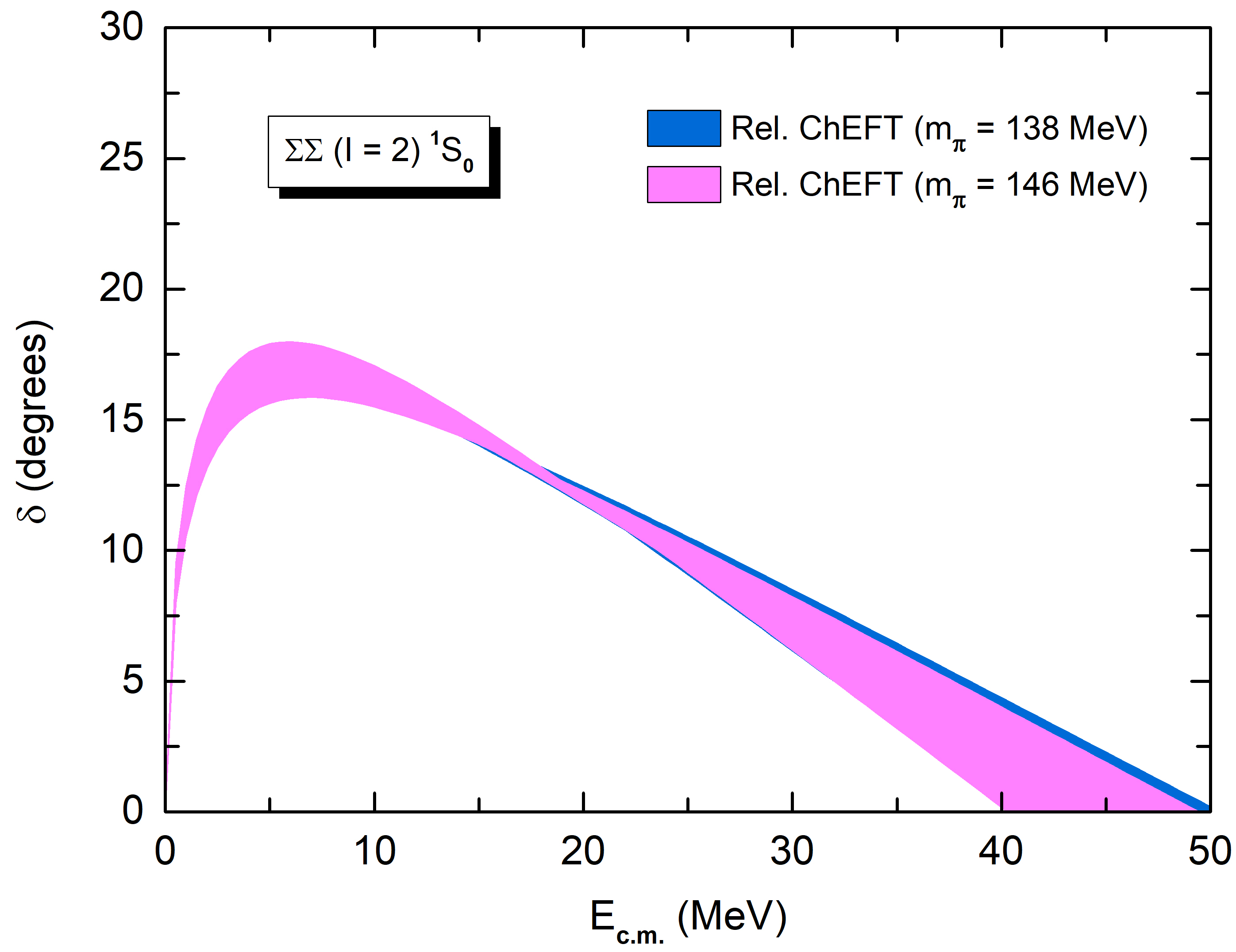}
  \caption{(color online) Prediction for the $\Sigma\Sigma$ ($I = 2$) $^1S_0$ phase shifts via SU(3) symmetry as a function of the c.m. kinetic energy $E_{\rm c.m.}$.
  }\label{Fig:PS_prediction}
\end{figure}

Due to the lack of hyperon scattering data, SU(3) flavor symmetry usually serves as a bridge to relate the YN and YY interactions with strangeness ranging from 0 to $-4$ in ChEFT and phenomenological models. It has been demonstrated that although the breaking of SU(3) flavor symmetry is non-negligible between different strangeness sectors, it holds approximately in the same strangeness sector~\cite{Haidenbauer2015EPJA51.17, KaiWen2018CPC42.014105, NPLQCD2021PRD103.054508}. Therefore, based on SU(3) flavor symmetry, we could determine the $\Sigma\Sigma~(I = 2)~^1S_0$ phase shifts from the above-mentioned lattice QCD simulations, as we have done in Sec.~\ref{subsection:fitting}. In Fig.~\ref{Fig:PS_prediction}, we predict an attractive $\Sigma\Sigma~(I = 2)~^1S_0$ interaction, but its strength is not strong enough to generate a bound state. In addition, the extrapolated phase shifts to the physical point are almost the same as those for the unphysical mass of $m_\pi = 146$ MeV. We note that the next-to-leading order heavy baryon (HB) ChEFT predicts similar but larger phase shifts (the maximum value of the phase shifts is about $30^{\circ}$)~\cite{Haidenbauer2015EPJA51.17, Haidenbauer2016NPA954.273}, which indicates a stronger attractive interaction in the $\Sigma\Sigma~(I = 2)~^1S_0$ channel.

\begin{figure}[h]
  \centering
  \includegraphics[width=0.6\textwidth]{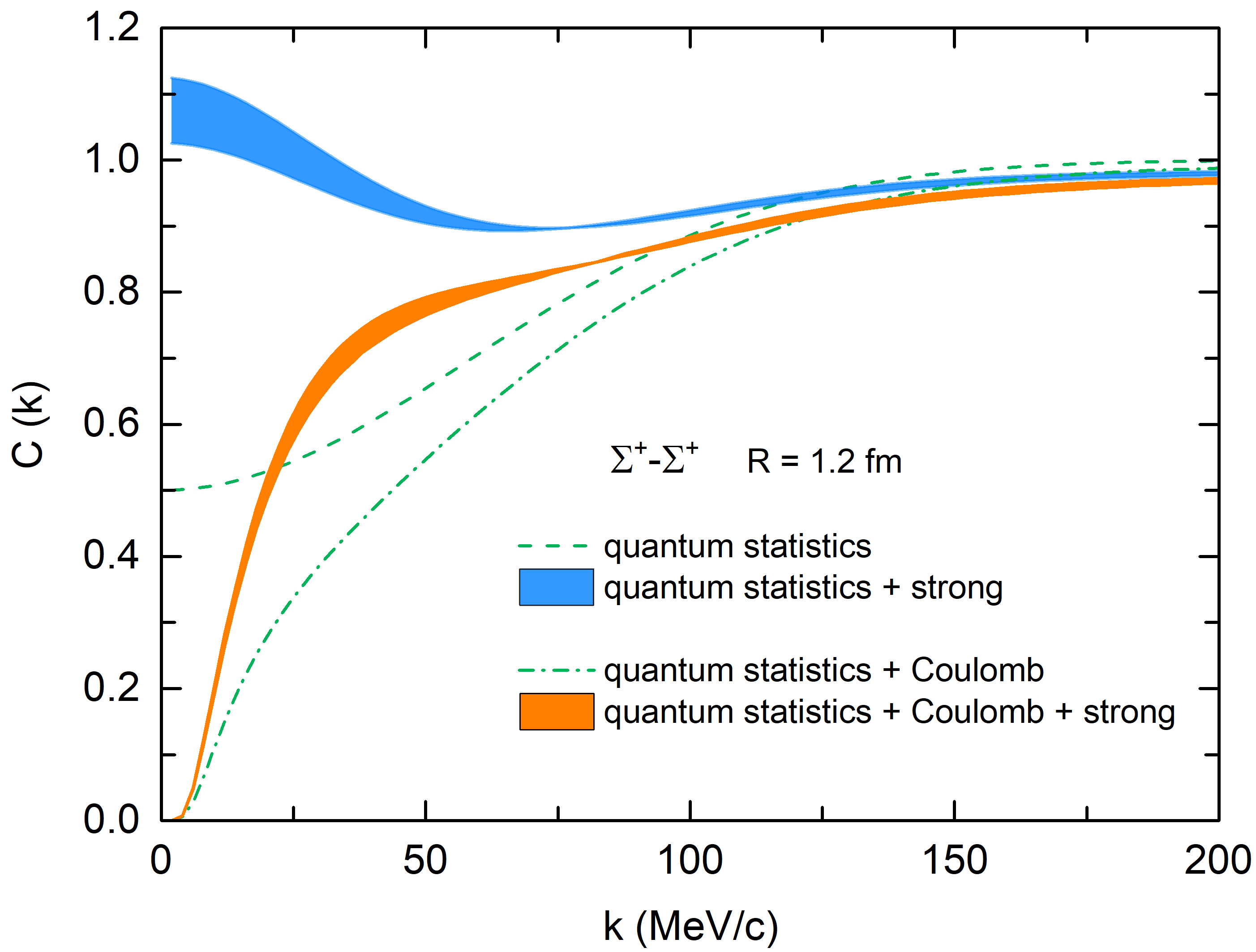}
  \caption{(color online) Predicted $\Sigma^+\Sigma^+$ correlation function as a function of the relative momentum $k$ in the covariant ChEFT. The orange (light blue) band denotes the result with (without) the Coulomb interaction taken into account. The results obtained with quantum statistical effects and with (without) the Coulomb interaction are also shown for comparison.
  }\label{Fig:CF_SigmapSigmap}
\end{figure}

For the $\Sigma^+\Sigma^+$ system, the relative wave function can be decomposed again according to  parity, namely,
\begin{subequations}
  \begin{align}
    \Psi_{SC,E}^{(-)}(\boldsymbol{r},\boldsymbol{k})&=\frac{\Psi_{SC}^{(-)}(\boldsymbol{r},\boldsymbol{k})+\Psi_{SC}^{(-)}(-\boldsymbol{r},\boldsymbol{k})}{\sqrt{2}}=\sqrt{2}\left[\phi_{sym}^C(\boldsymbol{r},\boldsymbol{k})-\phi_0^C(kr)+\psi_0^{SC}(r,k)\right],\label{Eq:WF_symmetry_Coulomb1}\\
    \Psi_{SC,O}^{(-)}(\boldsymbol{r},\boldsymbol{k})&=\frac{\Psi_{SC}^{(-)}(\boldsymbol{r},\boldsymbol{k})-\Psi_{SC}^{(-)}(-\boldsymbol{r},\boldsymbol{k})}{\sqrt{2}}=\sqrt{2}~\phi_{asym}^C(\boldsymbol{r},\boldsymbol{k}),\label{Eq:WF_symmetry_Coulomb2}
  \end{align}
\end{subequations}
where $\phi_{sym}^C$ and $\phi_{asym}^C$ denote the symmetrized and antisymmetrized  Coulomb wave function, respectively. Because of charge symmetry, the correlation functions of $\Sigma^+\Sigma^+$ and $\Sigma^-\Sigma^-$ are practically identical when the mass difference of baryons is neglected.. Substituting the relative wave functions \eqref{Eq:WF_symmetry_Coulomb1} and \eqref{Eq:WF_symmetry_Coulomb2} into the KP formula with the appropriate spin weights, the $\Sigma^+\Sigma^+$ correlation function reads
\begin{align}\label{Eq:CF_symmetry_Coulomb}
  C_{\Sigma^+\Sigma^+}(k)\simeq\frac{1}{2}\int{\rm d}\boldsymbol{r}~S_{12}(r)~\left[\left|\phi_{sym}^C(\boldsymbol{r},\boldsymbol{k})\right|^2+3\left|\phi_{asym}^C(\boldsymbol{r},\boldsymbol{k})\right|^2\right]+\frac{1}{2}\int_0^\infty4\pi r^2{\rm d}r~S_{12}(r)~\left[|\psi_0^{SC}(r,k)|^2-|\phi_0^C(kr)|^2\right],
\end{align}
where the first term on the right-hand side suppresses the $\Sigma^+\Sigma^+$ correlation over the whole $k$ range due to the quantum statistical effects and Coulomb repulsion. If the Coulomb interaction is neglected, the above expression can be reduced to the neutral case similar to Eq.~\eqref{Eq:CF_symmetry}. We note that only in Ref.~\cite{Haidenbauer2019NPA981.1}  the $\Sigma^+\Sigma^+$ correction function was studied, but the Coulomb interaction was neglected. On the other hand, in the present work, the covariant chiral potential, the Coulomb interaction, and the quantum statistical effects are fully taken into account. As shown in Fig.~\ref{Fig:CF_SigmapSigmap}, there is an enhancement in the full $C_{\Sigma^+\Sigma^+}$ below 100 MeV$/c$ compared to the case of pure quantum statistical effects or  Coulomb interaction. The enhancement is similar to the standard proton-proton correlation function, but there is no sharp peak near $20$ MeV$/c$, which indicates the breaking of SU(3) flavor symmetry. To facilitate comparison with other theoretical results, we also show the results obtained without the Coulomb interaction, shown by the light blue band in Fig.~\ref{Fig:CF_SigmapSigmap}. Compared to the HB ChEFT results~\cite{Haidenbauer2019NPA981.1}, our predictions show a weaker correlation in the low-momentum region, which can be traced back to the less attractive $\Sigma\Sigma$ interaction in the covariant ChEFT.

\subsection{Predictions for the $\Sigma^+\Lambda$ and $\Sigma^+\Sigma^-$ correlation functions}\label{subsection:SigmapLambda_SigmapSigmam}

For completeness, with the covariant chiral baryon-baryon interactions, we also predict the $\Sigma^+\Lambda$ and $\Sigma^+\Sigma^-$ correlation functions for the first time. Because of charge symmetry, the $\Sigma^+\Lambda$, $\Sigma^-\Lambda$, and $\Sigma^0\Lambda$ correlation functions are practically identical when the mass difference of baryons and the inelastic coupled-channel effects are neglected. Here, Eq.~\eqref{Eq:CF} and Eq.~\eqref{Eq:CF_Coulomb} are used to calculate the $\Sigma^+\Lambda$ and $\Sigma^+\Sigma^-$ correlation functions, respectively. Eq.~\eqref{Eq:Coupling} is used to evaluate the coupled-channel effects. It is worth noting that $\Sigma^+\Lambda$ in the spin-singlet channel can couple to $\Xi^0p$, while in the spin-triplet channel it can couple to $\Xi^0p$ and $\Sigma^0\Sigma^+$; $\Sigma^+\Sigma^-$ in the spin-singlet channel can couple to $\Lambda\Lambda$, $\Xi^0n$, $\Xi^-p$, and $\Sigma^0\Sigma^0$, while in the spin-triplet channel it can couple to $\Xi^0n$, $\Xi^-p$, and $\Sigma^0\Lambda$.

\begin{figure}[h]
  \centering
  \includegraphics[width=0.6\textwidth]{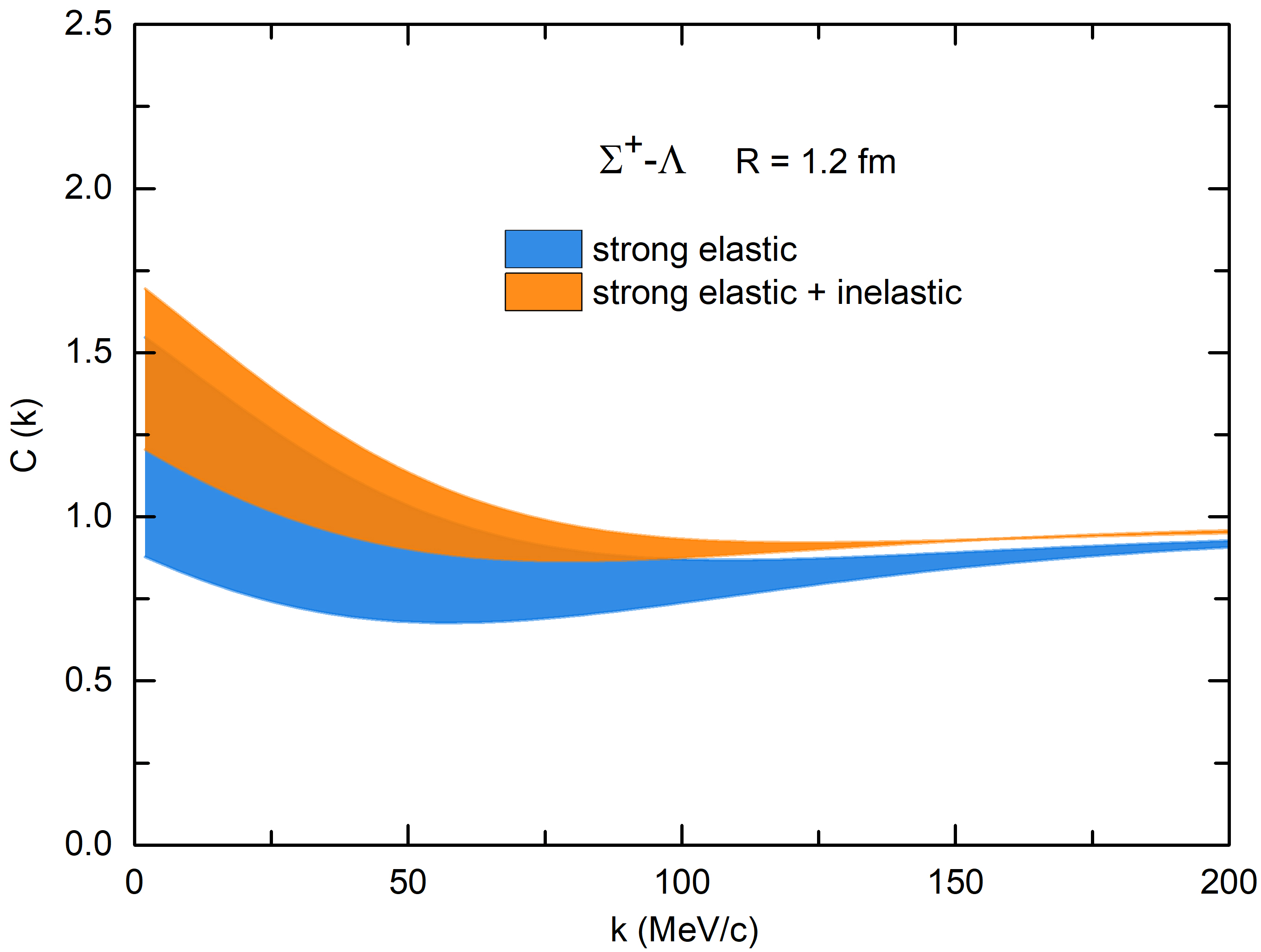}
  \caption{(color online) Predicted $\Sigma^+\Lambda$ correlation function as a function of the relative momentum $k$ in the covariant ChEFT. The orange (light blue) band denotes the result with (without) the inelastic coupled-channel effects taken into account. The bands reflect the variation of the cutoff in the range of $\Lambda_F$ = 550 – 700 MeV.
  }\label{Fig:CF_SigmapLambda}
\end{figure}

In Fig.~\ref{Fig:CF_SigmapLambda}, we show the $\Sigma^+\Lambda$ correlation functions with and without the inelastic coupled-channel effects (the orange and light blue bands, respectively). The enhancement in the light-blue curve at small momentum indicates the attractive $\Sigma^+\Lambda$ strong interaction. However, due to lack of direct constraints on the  $\Sigma\Lambda$ interaction, the predicted $\Sigma^+\Lambda$ correlation function has large uncertainties. Moreover, we note that the inelastic coupled-channel effects are significant, which can be traced back to the strong coupling between $\Xi^0p-$ and $\Sigma^+\Lambda$ in the $^3S_1$ channel.

\begin{figure}[h]
  \centering
  \includegraphics[width=0.6\textwidth]{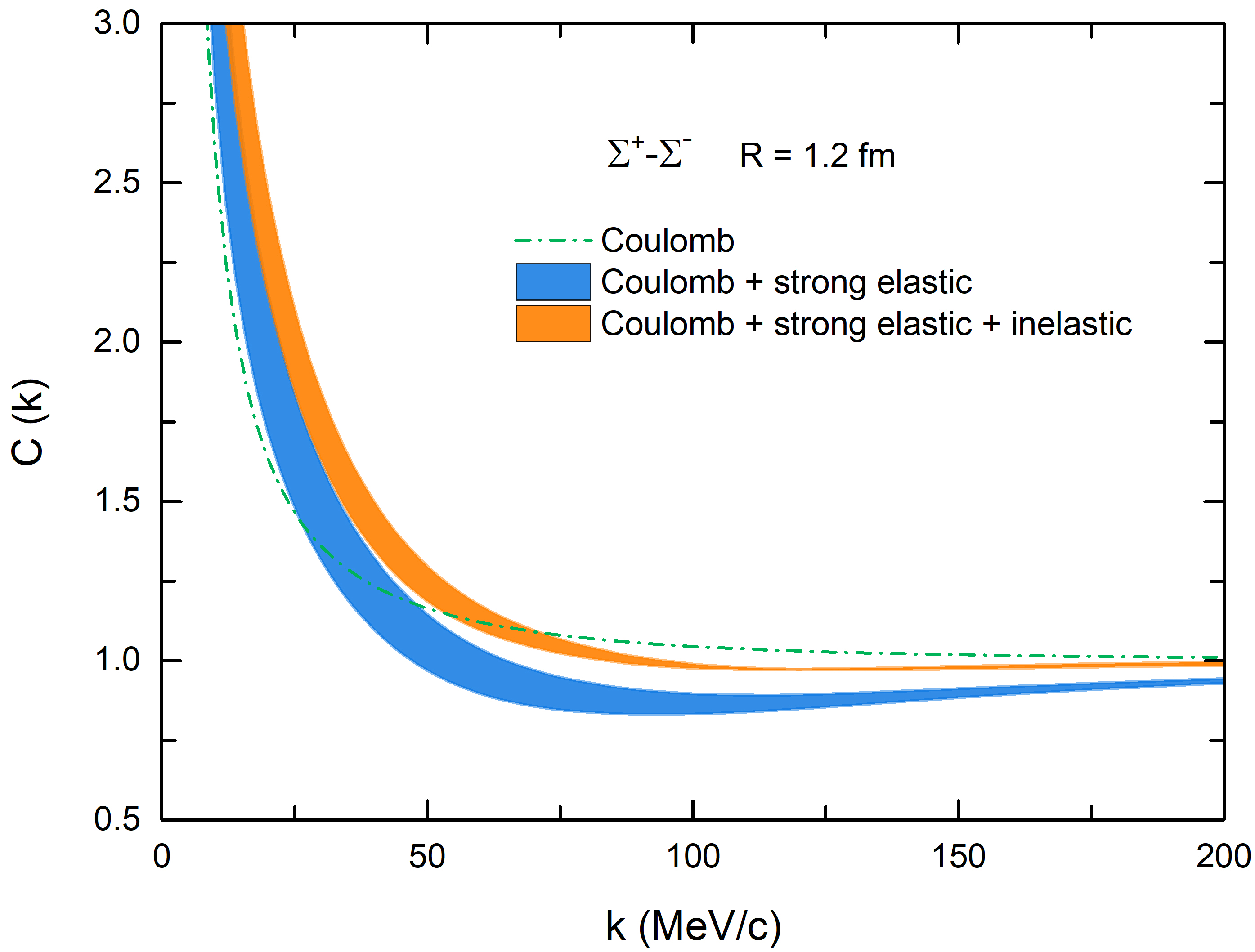}
  \caption{(color online) Predicted $\Sigma^+\Sigma^-$ correlation function as a function of the relative momentum $k$ in the covariant ChEFT. The orange (light blue) band denotes the result obtained with (without) the inelastic coupled-channel effects. The bands reflect the variation of the cutoff in the range of $\Lambda_F$ = 550 – 700 MeV. The result obtained with the Coulomb interaction are also shown for comparison.
  }\label{Fig:SigmapSigmam}
\end{figure}

The predicted $\Sigma^+\Sigma^-$ correlation functions are shown in Fig.~\ref{Fig:SigmapSigmam}. For this  system, the Coulomb interaction is attractive and its impact on the $\Sigma^+\Sigma^-$ correlation function is illustrated by the green dash-dotted line. It is seen that there is an enhancement in the light-blue curve at small momentum over the pure Coulomb case, while there is a depletion in the momentum region between 50 and 200 MeV$/c$. Nevertheless, we did not find any structure near the $\Sigma\Sigma$ threshold. We also note a significant inelastic coupled-channel contribution to the $\Sigma^+\Sigma^-$ correlation function. In fact, the inelastic contributions are mainly from the $\Sigma^0\Sigma^0-\Sigma^+\Sigma^-$ transition in the $^1S_0$ channel and $\Sigma^0\Lambda-\Sigma^+\Sigma^-$ transition in the $^3S_1$ channel. All these predictions could be tested by the ALICE Collaboration in the future.

\subsection{Source shape and size dependence}\label{subsection:Source_shape_size_dependence}

In our above study, following the experimental works~\cite{ALICE2019PLB797.134822, ALICE2019PRL123.112002, ALICE2020Nature588.232}, we have assumed a Gaussian source function and taken the size determined by experiments. In principle, however, the source of the hadron pair can be of other shapes, such as the Cauchy distribution discussed in Ref.~\cite{Oliver2017}. In the following, we study these two assumptions (source shape and size) in detail.

The Cauchy source function can be written in spherical coordinates as
\begin{align}\label{Eq:Cauchy}
    S_{12}(r,\theta,\varphi)=\left(\frac{R}{\pi}\right)^3\frac{r^2\sin\theta}{(r\sin\theta\cos\varphi)^2+R^2}\frac{1}{(r\sin\theta\sin\varphi)^2+R^2}\frac{1}{(r\cos\theta)^2+R^2},
\end{align}
where the numerator $r^2\sin\theta$ is the absolute value of the Jacobian determinant. Different from the Gaussian source, the Cauchy source has a much wider radial distribution, and emits particles in certain preferred directions, e.g., polar angle $\theta = \pi/2$ and azimuth angle $\varphi = \pi/2, \pi, 3\pi/2, 2\pi$. More details can be found in Ref.~\cite{Oliver2017}. Ultimately, the exact form of the source function can only be determined experimentally.

In the following, we recalculate the complete $\Lambda\Lambda$, $\Xi^-p$, $\Sigma^+\Sigma^+$, $\Sigma^+\Lambda$, and $\Sigma^+\Sigma^-$ correlation functions by using the Gaussian and Cauchy sources with source sizes $R = 0.9, 1.2, 1.5$ and $1.8$ fm. 

 \begin{table}[htbp]
  \centering
  \caption{Non-femtoscopic parameters $a$ and $b$ (in unite of $10^{-4}{\rm [MeV/c]^{-1}}$), and $\chi^2$/(d.o.f.) obtained by fitting to the ALICE data~\cite{ALICE2019PLB797.134822, ALICE2019PRL123.112002, ALICE2020Nature588.232} assuming different source shape and  size $R$ (in unite of fm) and with the  pair purity probability $\lambda$ suggested by experiments.  The calculations are performed with the covariant chiral baryon-baryon interactions ($\Lambda_F = 600$ MeV). See text for details.}
    \label{Tab:fitting}
    \setlength{\tabcolsep}{9pt}
    \begin{tabular}{c|c|c|c|c|r|c|c|c|c}
    \hline
    \hline
    \textbf{Collision} & \textbf{Source shap} & \textbf{R} & \textbf{$\lambda_{\Lambda\Lambda}$} & \textbf{$a_{\Lambda\Lambda}$} & \textbf{$b_{\Lambda\Lambda}$} & \textbf{$\lambda_{\Xi^-p}$} & \textbf{$a_{\Xi^-p}$} & \textbf{$b_{\Xi^-p}$} & \textbf{$\chi^2$/(d.o.f.)} \\
    \hline
    \multirow{8}[0]{*}{\textbf{$p-p\quad(13{\rm TeV})$}} & \multirow{4}[0]{*}{\textbf{Gaussian}} & 0.9   & \multirow{8}[0]{*}{0.338~\cite{ALICE2019PLB797.134822}} & 0.97 & 1.08 & \multirow{8}[0]{*}{1~\cite{ALICE2020Nature588.232}} & \multirow{8}[0]{*}{1~\cite{ALICE2020Nature588.232}} & \multirow{8}[0]{*}{0~\cite{ALICE2020Nature588.232}} & 2.18 \\
\cline{3-3}\cline{5-6}\cline{10-10}          &       & 1.2   &       & 0.95 & 1.39 &       &       &       & 0.98 \\
\cline{3-3}\cline{5-6}\cline{10-10}          &       & 1.5   &       & 0.95 & 1.51 &       &       &       & 1.00 \\
\cline{3-3}\cline{5-6}\cline{10-10}          &       & 1.8   &       & 0.94 & 1.57 &       &       &       & 1.27 \\
\cline{2-3}\cline{5-6}\cline{10-10}          & \multirow{4}[0]{*}{\textbf{Cauchy}} & 0.9   &       & 0.99 & 0.62 &       &       &       & 1.11 \\
\cline{3-3}\cline{5-6}\cline{10-10}          &       & 1.2   &       & 0.97 & 0.97 &       &       &       & 0.96 \\
\cline{3-3}\cline{5-6}\cline{10-10}          &       & 1.5   &       & 0.96 & 1.19 &       &       &       & 1.18 \\
\cline{3-3}\cline{5-6}\cline{10-10}          &       & 1.8   &       & 0.95 & 1.32 &       &       &       & 1.45 \\
    \hline
    \multirow{8}[0]{*}{\textbf{$p-{\rm Pb}\quad(5.02{\rm TeV})$}} & \multirow{4}[0]{*}{\textbf{Gaussian}} & 0.9   & \multirow{8}[0]{*}{0.239~\cite{ALICE2019PLB797.134822}} & 1.00 & 0.03 & \multirow{8}[0]{*}{0.513~\cite{ALICE2019PRL123.112002}} & 1.04 & -1.10 & 1.24 \\
\cline{3-3}\cline{5-6}\cline{8-10}          &       & 1.2   &       & 0.99 & 0.26 &       & 1.07 & -1.94 & 1.10 \\
\cline{3-3}\cline{5-6}\cline{8-10}          &       & 1.5   &       & 0.99 & 0.35 &       & 1.08 & -2.33 & 1.19 \\
\cline{3-3}\cline{5-6}\cline{8-10}          &       & 1.8   &       & 0.98 & 0.39 &       & 1.09 & -2.52 & 1.30 \\
\cline{2-3}\cline{5-6}\cline{8-10}          & \multirow{4}[0]{*}{\textbf{Cauchy}} & 0.9   &       & 1.02 & -0.29 &       & 1.06 & -1.51 & 1.05 \\
\cline{3-3}\cline{5-6}\cline{8-10}          &       & 1.2   &       & 1.01 & -0.05 &       & 1.07 & -2.09 & 1.08 \\
\cline{3-3}\cline{5-6}\cline{8-10}          &       & 1.5   &       & 1.00 & 0.11 &       & 1.08 & -2.37 & 1.18 \\
\cline{3-3}\cline{5-6}\cline{8-10}          &       & 1.8   &       & 0.99 & 0.21 &       & 1.09 & -2.54 & 1.29 \\
    \hline
    \hline
    \end{tabular}
  \label{tab:addlabel}
\end{table}

 \begin{figure}[h]
  \centering
  \includegraphics[width=0.95\textwidth]{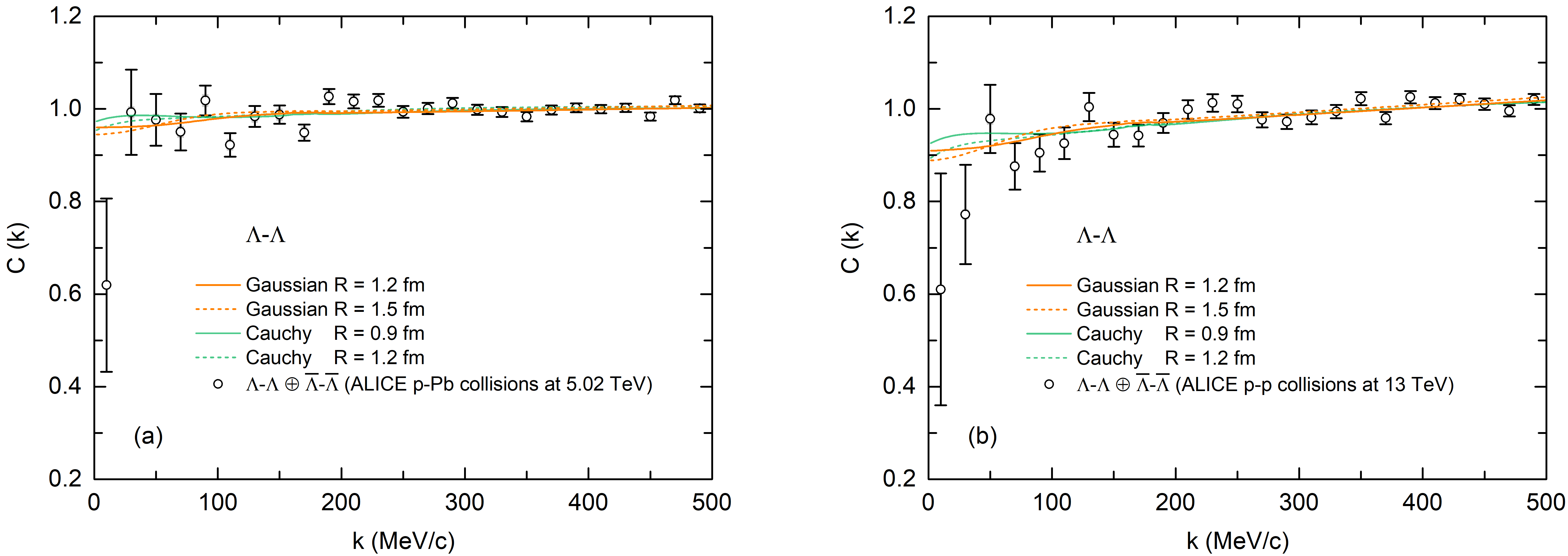}
  \caption{(color online) Theoretical $\Lambda\Lambda$ correlation function obtained with different source functions as a function of the relative momentum $k$, in comparison with the experimental data taken from $p$–Pb collisions at $\sqrt{s} = 5.02$ TeV \cite{ALICE2019PLB797.134822} and $p$–$p$ collisions at $\sqrt{s} = 13$ TeV \cite{ALICE2019PLB797.134822}. The theoretical results are calculated with the covariant chiral baryon-baryon interactions ($\Lambda_F = 600$ MeV).
  }\label{Fig:CF_LambdaLambda_Exp_Source}
\end{figure}

\begin{figure}[h]
  \centering
  \includegraphics[width=0.95\textwidth]{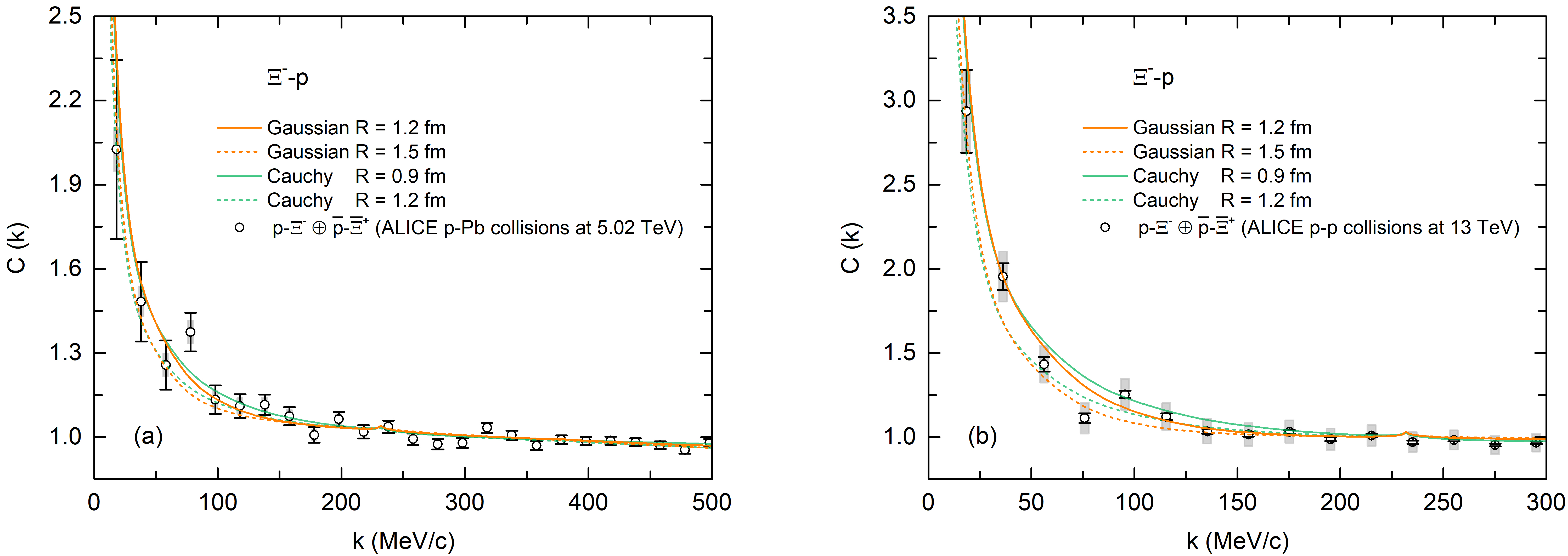}
  \caption{(color online) Theoretical $\Xi^-p$ correlation function obtained with different source functions as a function of the relative momentum $k$, in comparison with the experimental data taken from $p$–Pb collisions at $\sqrt{s} = 5.02$ TeV \cite{ALICE2019PRL123.112002} and $p$–$p$ collisions at $\sqrt{s} = 13$ TeV \cite{ALICE2020Nature588.232}. The theoretical results are calculated with the covariant chiral baryon-baryon interactions ($\Lambda_F = 600$ MeV).
  }\label{Fig:CF_Ximp_Exp_Source}
\end{figure}

Although remarkable agreements between the theoretical results and the experimental measurements have been found in Sec.~\ref{subsection:experiment}, we  recognize that in the $\Lambda\Lambda$ and $\Xi^-p$ femtoscopic analysis, accepting a Gaussian  source function and using the source radius determined from proton-proton correlations is only an assumption. It is necessary to explore  other possibilities, e.g., analyzing the current experimental data by using Gaussian and Cauchy sources with different sizes. In addition, to compare with the experimental data more reliably, we need to take care of the non-femtoscopic background term $(a+bk)$ in Eq.~\ref{Eq:CF_fit}. After fixing the source shape and size $R$, and using the pair purity probability $\lambda$ suggested by experiments~\cite{ALICE2019PLB797.134822, ALICE2019PRL123.112002, ALICE2020Nature588.232}, the non-femtoscopic parameters $a_{\Lambda\Lambda}$ and $b_{\Lambda\Lambda}$, and $a_{\Xi^-p}$ and $b_{\Xi^-p}$ in different collision systems are determined by fitting to the experimental  $\Lambda\Lambda$ and $\Xi^-p$ correlations simultaneously. The fitting results and corresponding $\chi^2$/(d.o.f.) are listed in Table~\ref{Tab:fitting}. It should be noticed that in obtaining the experimental $\Xi^-p$ correlation function of $p$-$p$ collisions the contaminations from particle mis-identification, the feed-down effect, and the non-femtoscopic effect have been subtracted. It is easily seen that for $p$–$p$ collisions  the $\chi^2$/(d.o.f.) is much scattered than that for  $p$–Pb collisions, which implies that the former  data are more sensitive to the source shape and size than the latter. Moreover, for $p$–$p$ collisions  the best $\chi^2$/(d.o.f.) are below 1 for both Gaussian and Cauchy sources, but for  $p$–Pb collisions the best $\chi^2$/(d.o.f.) are around 1.1 for both Gaussian and Cauchy sources. There is no obvious difference between the best fitting results  obtained with Gaussian and Cauchy sources for both collisions. 

The $\Lambda\Lambda$ and $\Xi^-p$ correlation functions obtained with different source functions  are shown in Figs.~\ref{Fig:CF_LambdaLambda_Exp_Source} and \ref{Fig:CF_Ximp_Exp_Source}. It is seen that due to the relatively large experimental uncertainties  in the low-momentum region, the curves obtained with different source functions cannot be well distinguished. The small difference between different theoretical curves can be partly attributed to  the small pair purity probability, especially for the $\Lambda\Lambda$ pair, which suppresses the sensitivity of the correlation function to the source shape and size. See below for more related discussions. 

\begin{figure}[h]
  \centering
  \includegraphics[width=0.95\textwidth]{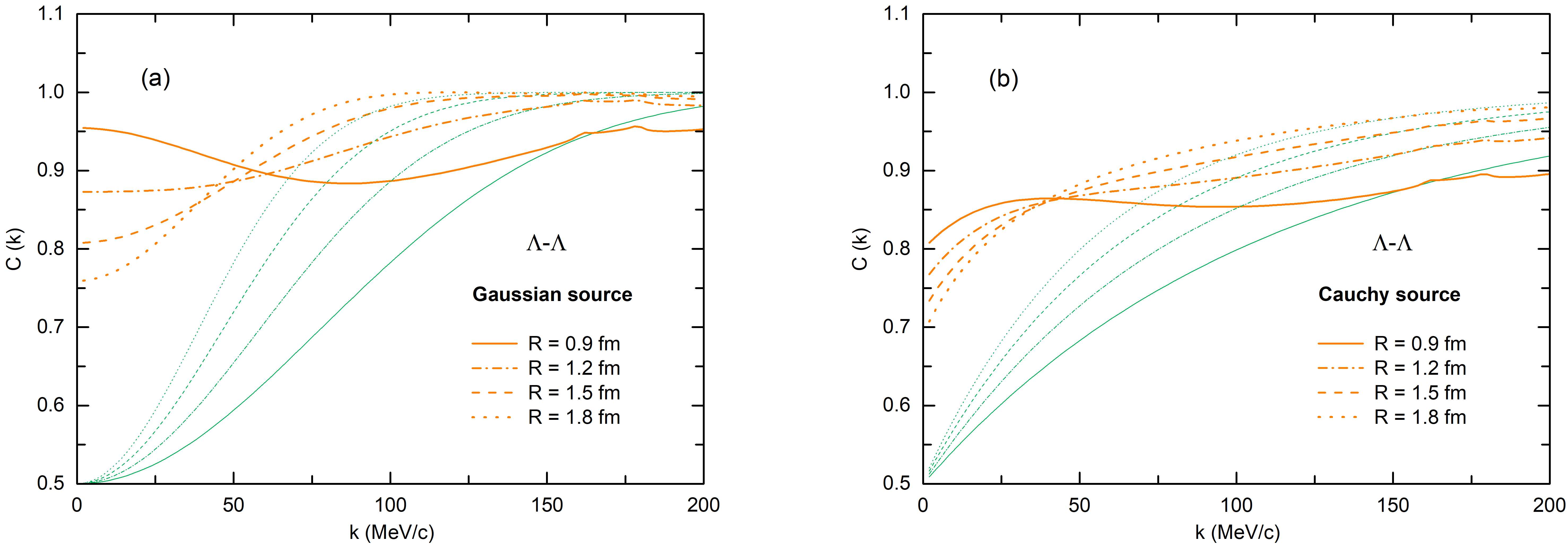}
  \caption{(color online) $\Lambda\Lambda$ correlation function as a function of the relative momentum $k$, calculated with the Gaussian (left) and Cauchy (right) sources and with source sizes $R = 0.9, 1.2, 1.5$ and $1.8$ fm. The orange (thick) lines denote the full results with the covariant chiral baryon-baryon interactions (obtained with $\Lambda_F = 600$ MeV). For comparison, the results obtained by considering only quantum statistical effects are shown  by green (thin) lines.
  }\label{Fig:CF_LambdaLambda_source}
\end{figure}

\begin{figure}[h]
  \centering
  \includegraphics[width=0.95\textwidth]{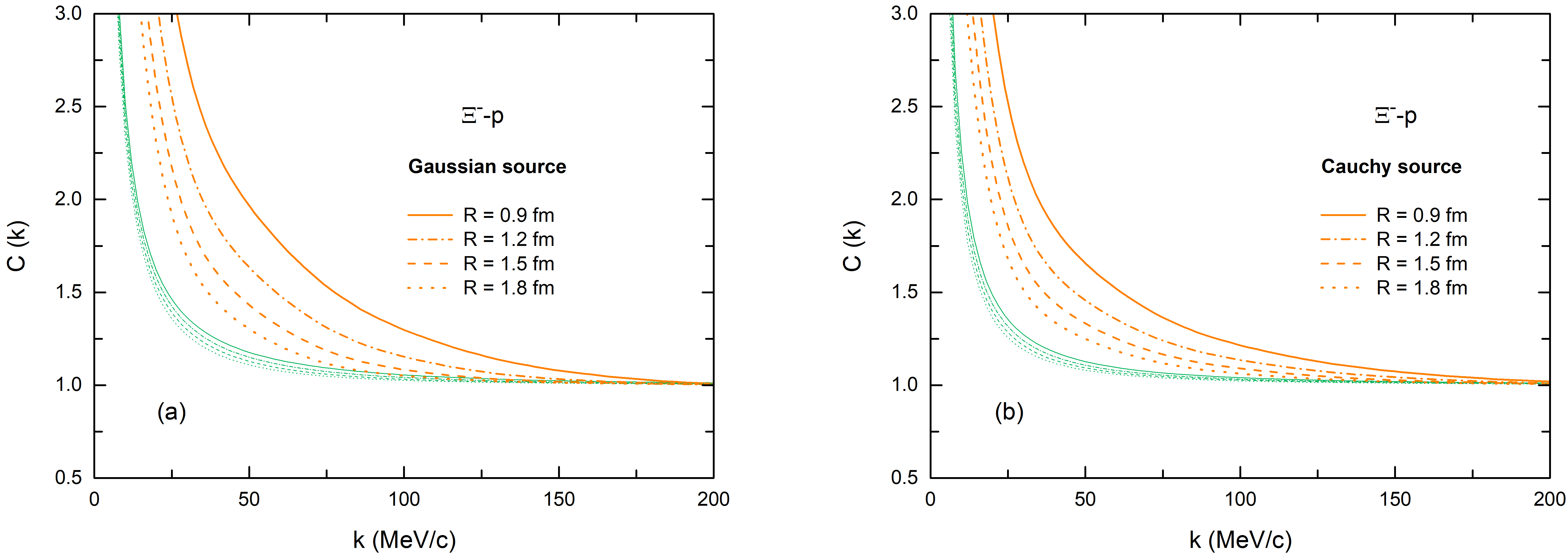}
  \caption{(color online) $\Xi^-p$ correlation function as a function of the relative momentum $k$, calculated with the Gaussian (left) and Cauchy (right) sources and with source sizes $R = 0.9, 1.2, 1.5$ and $1.8$ fm. The orange (thick) lines denote the full results with the covariant chiral baryon-baryon interactions (obtained with $\Lambda_F = 600$ MeV). For comparison, the results obtained by considering only the Coulomb contribution are shown by green (thin) lines.
  }\label{Fig:CF_Ximp_source}
\end{figure}

To better understand the source shape and size dependence, we  neglect the experimental corrections by setting $a=1$, $b=0$, and $\lambda=1$ and focus on the impact of the source function. Fig.~\ref{Fig:CF_LambdaLambda_source} shows the source shape and size dependence of the $\Lambda\Lambda$ correlation function. The correlation functions obtained with Gaussian and Cauchy sources are quite different, especially in the low-momentum region. For the same source size $R$, the result obtained with the Cauchy source is smaller than that obtained with the Gaussian source. We  note that the momentum evolution  of the $\Lambda\Lambda$ correlation function as function of the source size $R$ is non-monotonic for both Gaussian and Cauchy sources. Moreover, the two cusp-like structures (around $k \approx 170$ MeV$/c$) become more pronounced with decreasing  $R$. The source shape and size dependence of the $\Xi^-p$ correlation function is displayed in Fig.~\ref{Fig:CF_Ximp_source}. The $\Xi^-p$ correlation functions calculated with the Cauchy source are qualitatively  similar to  the Gaussian results, but quantitatively smaller for the same size and momentum. We note that the enhancement of the $\Xi^-p$ correlation function due to the strong interaction becomes larger as the  source size becomes smaller.
 
\begin{figure}[h]
  \centering
  \includegraphics[width=0.95\textwidth]{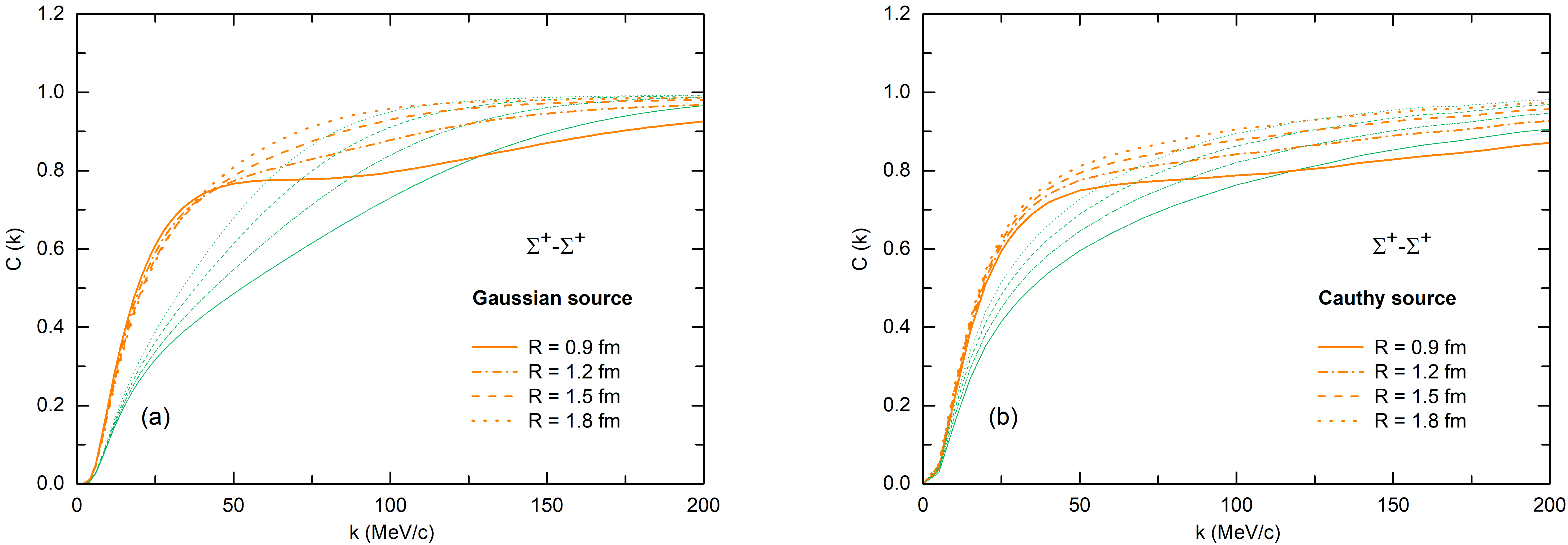}
  \caption{(color online) $\Sigma^+\Lambda$ correlation function as a function of the relative momentum $k$, calculated with the Gaussian (left) and Cauchy (right) sources and with source sizes $R = 0.9, 1.2, 1.5$ and $1.8$ fm. The orange (thick) lines denote the full results with the covariant chiral baryon-baryon interactions (obtained with $\Lambda_F = 600$ MeV). For comparison, the results obtained by considering quantum statistical effects and the Coulomb interaction are shown by green (thin) lines.
  }\label{Fig:CF_SigmapSigmap_source}
\end{figure}

\begin{figure}[h]
  \centering
  \includegraphics[width=0.95\textwidth]{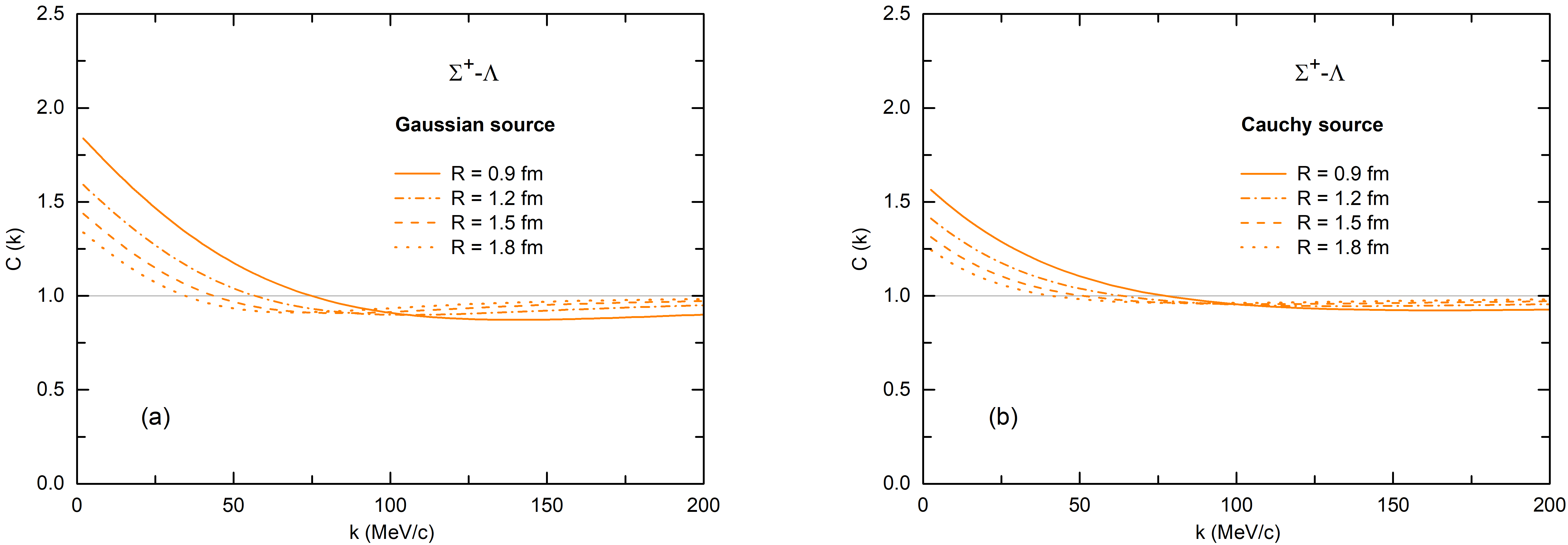}
  \caption{(color online) $\Sigma^+\Lambda$ correlation function as a function of the relative momentum $k$, calculated with the Gaussian (left) and Cauchy (right) sources and with source sizes $R = 0.9, 1.2, 1.5$ and $1.8$fm. The orange (thick) lines denote the full results with the covariant chiral baryon-baryon interactions (obtained with $\Lambda_F = 600$ MeV).
  }\label{Fig:CF_SigmapLambda_source}
\end{figure}

\begin{figure}[h]
  \centering
  \includegraphics[width=0.95\textwidth]{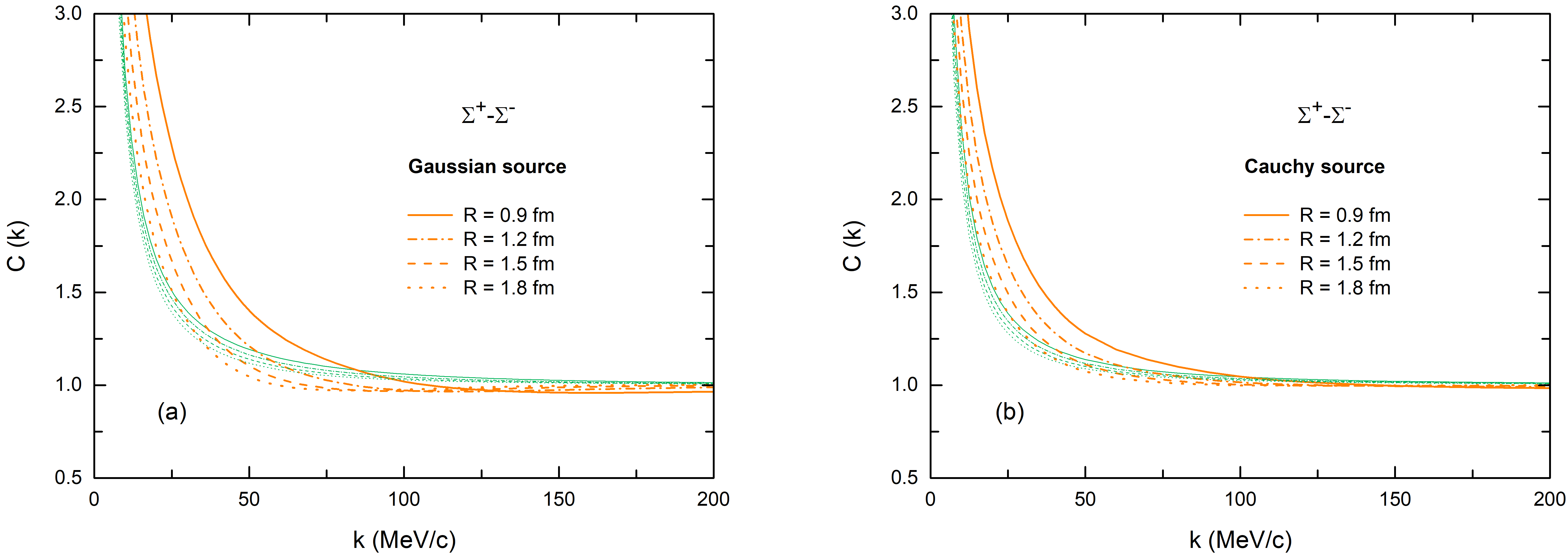}
  \caption{(color online) $\Sigma^+\Sigma^-$ correlation function as a function of the relative momentum $k$, calculated with the Gaussian (left) and Cauchy (right) sources and with source sizes $R = 0.9, 1.2, 1.5$ and $1.8$ fm. The orange (thick) lines denote the full results with the covariant chiral baryon-baryon interactions (obtained with $\Lambda_F = 600$ MeV). For comparison, the results obtained by considering only the Coulomb contribution are shown by green (thin) lines.
  }\label{Fig:CF_SigmapSigmam_source}
\end{figure}

The source shape and size dependence of the $\Sigma^+\Sigma^+$, $\Sigma^+\Lambda$, and $\Sigma^+\Sigma^-$ correlation functions are predicted in Figs. \ref{Fig:CF_SigmapSigmap_source}, \ref{Fig:CF_SigmapLambda_source}, and \ref{Fig:CF_SigmapSigmam_source}, respectively. With respect to the source shape dependence, although these three correlation functions obtained with the Cauchy source are qualitatively similar to the corresponding Gaussian results, there are still visible differences. For example, for the same source size $R$, the $\Sigma^+\Sigma^+$ correlation function obtained with the Cauchy source is smaller than the Gaussian one in the momentum region between 50 and 200 MeV$/c$; the $\Sigma^+\Lambda$ and $\Sigma^+\Sigma^-$ correlation functions obtained with the Cauchy source are both smaller than the corresponding Gaussian results in the low-momentum region. With respect to the source size dependence, the enhancements of $\Sigma^+\Sigma^+$, $\Sigma^+\Lambda$, and $\Sigma^+\Sigma^-$ correlation functions due to the strong interactions are all dramatic for $R = 0.9$ fm but become smaller for larger source sizes. This is understandable because the strong interaction is  of short-range nature. It is also interesting to note that the $\Sigma^+\Sigma^+$ correlation function is not sensitive to the source shape or size in the low-momentum region between 0 and 50 MeV$/c$, where the Coulomb repulsion plays a dominant role.

\section{Summary and Outlook}

In this work, we studied the strangeness $S = -2$ baryon-baryon interactions and the corresponding momentum correlation functions in the leading order covariant SU(3) ChEFT. Considering all the coupled channels, we first determined the hyperon-nucleon and hyperon-hyperon $S$-wave interactions by fitting to the state-of-the-art lattice QCD simulations. With the so-obtained strong interactions and considering quantum statistical effects, the Coulomb interaction, and all the coupled channels, we computed the $\Lambda\Lambda$ and $\Xi^-p$ correlation functions with a spherical Gaussian source. We found that there are significant enhancements in both systems due to the strong interaction. The numerical results showed that the inelastic $\Xi^0n$ and $\Xi^-p$ channels led to appreciable cusp-like structures in the $\Lambda\Lambda$ correlation function around the relative momentum $k \approx 170$ MeV$/c$, and the opening of the inelastic $\Sigma^0\Lambda$ channel also left a trace in the $\Xi^-p$ correlation function around $k \approx 230$ MeV$/c$, which are expected to be observed with high precision correlation techniques in future experiments. After using the source radius suggested by proton-proton correlations and reasonable ratios of weights $\omega_\beta/\omega_\alpha$, we compared the theoretical correlation functions with the recent experimental data. The agreement between theoretical descriptions and experimental measurements turned out to be very good, demonstrating the reliability of the covariant chiral $S = -2$ baryon-baryon interactions. Moreover, we predicted an attractive $\Sigma\Sigma~(I = 2)~^1S_0$ interaction, but the strength was not strong enough to generate a bound state. We further predicted the $\Sigma^+\Sigma^+$, $\Sigma^+\Lambda$, and $\Sigma^+\Sigma^-$ correlation functions, which can be tested by future experiment. Finally, we investigated the influence of the source shape and size on the correlation functions by comparing the Gaussian source with the Cauchy source and varying the source size. We showed that the current experimental data are not sensitive to the source shape. Hopefully future precise data can address this issue in more detail.

With the upgraded ALICE apparatus and the larger data sample size expected~ \cite{ALICE-PUBLIC-2020-005}, more interactions in the $S = -2$ sector (such as $\Sigma\Lambda$ and $\Sigma\Sigma$) could be measured in the upcoming LHC Run 3 and Run 4. In addition, studies of hypernuclei provide an promising alternative to extract information on the $S = -2$ baryon-baryon interactions from the photon-induced reactions at JLab or $K^-$-induced reactions at J-PARC~\cite{Miller2013CJP51.466, CLAS2021PRL127.272303, Ahn2011JPARC}. These studies will substantially advance our knowledge about the YN and YY interactions in the not-so-far future.

\section*{Acknowledgements}
The authors thank Johann Haidenbauer for useful discussions concerning the Coulomb interaction and Qiang Zhao for the suggestion to study the $\Sigma^+\Sigma^-$ correlation function. This work is partly supported by the National Natural Science Foundation of China under Grant Nos.11735003, 11975041, and 11961141004.

\bibliographystyle{elsarticle-num}
\bibliography{BB.bib}

\begin{thebibliography}{100}
\expandafter\ifx\csname url\endcsname\relax
  \def\url#1{\texttt{#1}}\fi
\expandafter\ifx\csname urlprefix\endcsname\relax\def\urlprefix{URL }\fi
\expandafter\ifx\csname href\endcsname\relax
  \def\href#1#2{#2} \def\path#1{#1}\fi

\bibitem{Jaffe1977PRL38.195}
R.~L. Jaffe, \href{https://link.aps.org/doi/10.1103/PhysRevLett.38.195}{Perhaps
  a stable dihyperon}, Phys. Rev. Lett. 38 (1977) 195--198.
\newblock \href {https://doi.org/10.1103/PhysRevLett.38.195}
  {\path{doi:10.1103/PhysRevLett.38.195}}.
\newline\urlprefix\url{https://link.aps.org/doi/10.1103/PhysRevLett.38.195}

\bibitem{NPLQCD2011PRL106.162001}
S.~R. Beane, E.~Chang, W.~Detmold, B.~Joo, H.~W. Lin, T.~C. Luu, K.~Orginos,
  A.~Parre\~no, M.~J. Savage, A.~Torok, A.~Walker-Loud,
  \href{https://link.aps.org/doi/10.1103/PhysRevLett.106.162001}{Evidence for a
  bound $\mathrm{H}$ dibaryon from lattice $\mathrm{QCD}$}, Phys. Rev. Lett.
  106 (2011) 162001.
\newblock \href {https://doi.org/10.1103/PhysRevLett.106.162001}
  {\path{doi:10.1103/PhysRevLett.106.162001}}.
\newline\urlprefix\url{https://link.aps.org/doi/10.1103/PhysRevLett.106.162001}

\bibitem{HALQCD2011PRL106.162002}
T.~Inoue, N.~Ishii, S.~Aoki, T.~Doi, T.~Hatsuda, Y.~Ikeda, K.~Murano,
  H.~Nemura, K.~Sasaki,
  \href{https://link.aps.org/doi/10.1103/PhysRevLett.106.162002}{Bound
  $\mathrm{H}$ dibaryon in flavor $\mathrm{SU(3)}$ limit of lattice
  $\mathrm{QCD}$}, Phys. Rev. Lett. 106 (2011) 162002.
\newblock \href {https://doi.org/10.1103/PhysRevLett.106.162002}
  {\path{doi:10.1103/PhysRevLett.106.162002}}.
\newline\urlprefix\url{https://link.aps.org/doi/10.1103/PhysRevLett.106.162002}

\bibitem{Shanahan2011PRL107.092004}
P.~E. Shanahan, A.~W. Thomas, R.~D. Young,
  \href{https://link.aps.org/doi/10.1103/PhysRevLett.107.092004}{Mass of the
  $\mathrm{H}$ dibaryon}, Phys. Rev. Lett. 107 (2011) 092004.
\newblock \href {https://doi.org/10.1103/PhysRevLett.107.092004}
  {\path{doi:10.1103/PhysRevLett.107.092004}}.
\newline\urlprefix\url{https://link.aps.org/doi/10.1103/PhysRevLett.107.092004}

\bibitem{Green2021PRL127.242003}
J.~R. Green, A.~D. Hanlon, P.~M. Junnarkar, H.~Wittig,
  \href{https://link.aps.org/doi/10.1103/PhysRevLett.127.242003}{Weakly bound
  $h$ dibaryon from su(3)-flavor-symmetric qcd}, Phys. Rev. Lett. 127 (2021)
  242003.
\newblock \href {https://doi.org/10.1103/PhysRevLett.127.242003}
  {\path{doi:10.1103/PhysRevLett.127.242003}}.
\newline\urlprefix\url{https://link.aps.org/doi/10.1103/PhysRevLett.127.242003}

\bibitem{Takahashi2001PRL87.212502}
H.~Takahashi, et~al.,
  \href{https://link.aps.org/doi/10.1103/PhysRevLett.87.212502}{Observation of
  a $_{\ensuremath{\Lambda}\ensuremath{\Lambda}}^{6}\mathrm{He}$ double
  hypernucleus}, Phys. Rev. Lett. 87 (2001) 212502.
\newblock \href {https://doi.org/10.1103/PhysRevLett.87.212502}
  {\path{doi:10.1103/PhysRevLett.87.212502}}.
\newline\urlprefix\url{https://link.aps.org/doi/10.1103/PhysRevLett.87.212502}

\bibitem{Hiyama2020PRL124.092501}
E.~Hiyama, K.~Sasaki, T.~Miyamoto, T.~Doi, T.~Hatsuda, Y.~Yamamoto, T.~A.
  Rijken,
  \href{https://link.aps.org/doi/10.1103/PhysRevLett.124.092501}{Possible
  lightest $\mathrm{\ensuremath{\Xi}}$ hypernucleus with modern
  $\mathrm{\ensuremath{\Xi}}\mathrm{N}$ interactions}, Phys. Rev. Lett. 124
  (2020) 092501.
\newblock \href {https://doi.org/10.1103/PhysRevLett.124.092501}
  {\path{doi:10.1103/PhysRevLett.124.092501}}.
\newline\urlprefix\url{https://link.aps.org/doi/10.1103/PhysRevLett.124.092501}

\bibitem{JPARC2021PRL126.062501}
S.~H. Hayakawa, et~al.,
  \href{https://link.aps.org/doi/10.1103/PhysRevLett.126.062501}{Observation of
  coulomb-assisted nuclear bound state of
  ${\mathrm{\ensuremath{\Xi}}}^{\ensuremath{-}}--^{14}\mathrm{N}$ system},
  Phys. Rev. Lett. 126 (2021) 062501.
\newblock \href {https://doi.org/10.1103/PhysRevLett.126.062501}
  {\path{doi:10.1103/PhysRevLett.126.062501}}.
\newline\urlprefix\url{https://link.aps.org/doi/10.1103/PhysRevLett.126.062501}

\bibitem{Lonardoni2014PRC89.014314}
D.~Lonardoni, F.~Pederiva, S.~Gandolfi,
  \href{https://link.aps.org/doi/10.1103/PhysRevC.89.014314}{Accurate
  determination of the interaction between $\ensuremath{\Lambda}$ hyperons and
  nucleons from auxiliary field diffusion monte carlo calculations}, Phys. Rev.
  C 89 (2014) 014314.
\newblock \href {https://doi.org/10.1103/PhysRevC.89.014314}
  {\path{doi:10.1103/PhysRevC.89.014314}}.
\newline\urlprefix\url{https://link.aps.org/doi/10.1103/PhysRevC.89.014314}

\bibitem{Maslov2015PLB748.369}
K.~A. Maslov, E.~E. Kolomeitsev, D.~N. Voskresensky,
  \href{https://www.sciencedirect.com/science/article/pii/S0370269315005420}{Solution
  of the hyperon puzzle within a relativistic mean-field model}, Phys. Lett. B
  748 (2015) 369--375.
\newblock \href
  {https://doi.org/https://doi.org/10.1016/j.physletb.2015.07.032}
  {\path{doi:https://doi.org/10.1016/j.physletb.2015.07.032}}.
\newline\urlprefix\url{https://www.sciencedirect.com/science/article/pii/S0370269315005420}

\bibitem{Oertel2015JPG42.075202}
M.~Oertel, C.~Provid{\^{e}}ncia, F.~Gulminelli, A.~R. Raduta,
  \href{https://doi.org/10.1088/0954-3899/42/7/075202}{Hyperons in neutron star
  matter within relativistic mean-field models}, J. Phys. G: Nucl. Part. Phys.
  42~(7) (2015) 075202.
\newblock \href {https://doi.org/10.1088/0954-3899/42/7/075202}
  {\path{doi:10.1088/0954-3899/42/7/075202}}.
\newline\urlprefix\url{https://doi.org/10.1088/0954-3899/42/7/075202}

\bibitem{Lim2015IJMPE24.1550100}
Y.~Lim, C.~H. Hyun, K.~Kwak, C.~H. Lee,
  \href{https://doi.org/10.1142/S0218301315501001}{Hyperon puzzle of neutron
  stars with skyrme force models}, Int. J. Mod. Phys. E 24 (2015) 1550100.
\newblock \href {https://doi.org/10.1142/S0218301315501001}
  {\path{doi:10.1142/S0218301315501001}}.
\newline\urlprefix\url{https://doi.org/10.1142/S0218301315501001}

\bibitem{Engelmann1966PL21.587}
R.~Engelmann, H.~Filthuth, V.~Hepp, E.~Kluge,
  \href{http://www.sciencedirect.com/science/article/pii/0031916366913102}{Inelastic
  $\ensuremath{\Sigma}$−$\mathrm{p}$ interactions at low momenta}, Phys.
  Lett. 21~(5) (1966) 587 -- 589.
\newblock \href {https://doi.org/10.1016/0031-9163(66)91310-2}
  {\path{doi:10.1016/0031-9163(66)91310-2}}.
\newline\urlprefix\url{http://www.sciencedirect.com/science/article/pii/0031916366913102}

\bibitem{Alexander1968PR173.1452}
G.~Alexander, U.~Karshon, A.~Shapira, G.~Yekutieli, R.~Engelmann, H.~Filthuth,
  W.~Lughofer, \href{https://link.aps.org/doi/10.1103/PhysRev.173.1452}{Study
  of the $\ensuremath{\Lambda}\ensuremath{-}\mathrm{N}$ system in low-energy
  $\ensuremath{\Lambda}\ensuremath{-}\mathrm{p}$ elastic scattering}, Phys.
  Rev. 173 (1968) 1452--1460.
\newblock \href {https://doi.org/10.1103/PhysRev.173.1452}
  {\path{doi:10.1103/PhysRev.173.1452}}.
\newline\urlprefix\url{https://link.aps.org/doi/10.1103/PhysRev.173.1452}

\bibitem{Sechi1968PR75.1735}
B.~Sechi-Zorn, B.~Kehoe, J.~Twitty, R.~A. Burnstein,
  \href{https://link.aps.org/doi/10.1103/PhysRev.175.1735}{Low-energy
  $\ensuremath{\Lambda}$-proton elastic scattering}, Phys. Rev. 175 (1968)
  1735--1740.
\newblock \href {https://doi.org/10.1103/PhysRev.175.1735}
  {\path{doi:10.1103/PhysRev.175.1735}}.
\newline\urlprefix\url{https://link.aps.org/doi/10.1103/PhysRev.175.1735}

\bibitem{Hepp1968ZP214.71}
V.~Hepp, H.~Schleich, \href{https://doi.org/10.1007/BF01380085}{A new
  determination of the capture ratio $\ensuremath{r_c = \frac{{\sum ^ - p \to
  \sum ^0 n}}{{(\sum ^ - p \to \sum ^0 n) + (\sum ^ - p \to \Lambda ^0 n)}}}$,
  the $\ensuremath{\Lambda^0}$-lifetime and the
  $\ensuremath{\Sigma^-}$-$\ensuremath{\Lambda^0}$ mass difference}, Z. Phys.
  214 (1968) 71.
\newblock \href {https://doi.org/10.1007/BF01380085}
  {\path{doi:10.1007/BF01380085}}.
\newline\urlprefix\url{https://doi.org/10.1007/BF01380085}

\bibitem{Eisele1971PLB37.204}
F.~Eisele, H.~Filthuth, W.~Föhlisch, V.~Hepp, G.~Zech,
  \href{http://www.sciencedirect.com/science/article/pii/0370269371900530}{Elastic
  $\ensuremath{\Sigma^\pm}$$\mathrm{p}$ scattering at low energies}, Phys.
  Lett. B 37~(2) (1971) 204 -- 206.
\newblock \href {https://doi.org/10.1016/0370-2693(71)90053-0}
  {\path{doi:10.1016/0370-2693(71)90053-0}}.
\newline\urlprefix\url{http://www.sciencedirect.com/science/article/pii/0370269371900530}

\bibitem{CLAS2021PRL127.272303}
J.~Rowley, et~al.,
  \href{https://link.aps.org/doi/10.1103/PhysRevLett.127.272303}{Improved
  $\mathrm{\ensuremath{\Lambda}}p$ elastic scattering cross sections between
  0.9 and $2.0\text{ }\text{ }\mathrm{GeV}/c$ as a main ingredient of the
  neutron star equation of state}, Phys. Rev. Lett. 127 (2021) 272303.
\newblock \href {https://doi.org/10.1103/PhysRevLett.127.272303}
  {\path{doi:10.1103/PhysRevLett.127.272303}}.
\newline\urlprefix\url{https://link.aps.org/doi/10.1103/PhysRevLett.127.272303}

\bibitem{Ahn2006PLB633.214}
J.~K. Ahn, et~al.,
  \href{http://www.sciencedirect.com/science/article/pii/S0370269305018770}{Measurement
  of the $\mathrm{\Xi}$−$\mathrm{p}$ scattering cross sections at low
  energy}, Phys. Lett. B 633~(2) (2006) 214 -- 218.
\newblock \href {https://doi.org/10.1016/j.physletb.2005.12.057}
  {\path{doi:10.1016/j.physletb.2005.12.057}}.
\newline\urlprefix\url{http://www.sciencedirect.com/science/article/pii/S0370269305018770}

\bibitem{Stoks1999PRC59.3009}
V.~G.~J. Stoks, T.~A. Rijken,
  \href{https://link.aps.org/doi/10.1103/PhysRevC.59.3009}{Soft-core
  baryon-baryon potentials for the complete baryon octet}, Phys. Rev. C 59
  (1999) 3009--3020.
\newblock \href {https://doi.org/10.1103/PhysRevC.59.3009}
  {\path{doi:10.1103/PhysRevC.59.3009}}.
\newline\urlprefix\url{https://link.aps.org/doi/10.1103/PhysRevC.59.3009}

\bibitem{Polinder2006NPA779.244}
H.~Polinder, J.~Haidenbauer, U.-G. Meißner,
  \href{http://www.sciencedirect.com/science/article/pii/S0375947406006312}{Hyperon–nucleon
  interactions—a chiral effective field theory approach}, Nucl. Phys. A 779
  (2006) 244 -- 266.
\newblock \href
  {https://doi.org/https://doi.org/10.1016/j.nuclphysa.2006.09.006}
  {\path{doi:https://doi.org/10.1016/j.nuclphysa.2006.09.006}}.
\newline\urlprefix\url{http://www.sciencedirect.com/science/article/pii/S0375947406006312}

\bibitem{Fujiwara2007PPNP58.439}
Y.~Fujiwara, Y.~Suzuki, C.~Nakamoto,
  \href{http://www.sciencedirect.com/science/article/pii/S0146641006000718}{Baryon–baryon
  interactions in the $\mathrm{SU6}$ quark model and their applications to
  light nuclear systems}, Prog. Part. Nucl. Phys. 58~(2) (2007) 439 -- 520.
\newblock \href {https://doi.org/https://doi.org/10.1016/j.ppnp.2006.08.001}
  {\path{doi:https://doi.org/10.1016/j.ppnp.2006.08.001}}.
\newline\urlprefix\url{http://www.sciencedirect.com/science/article/pii/S0146641006000718}

\bibitem{Haidenbauer2010PLB684.275}
J.~Haidenbauer, U.-G. Meißner,
  \href{http://www.sciencedirect.com/science/article/pii/S0370269310000730}{Predictions
  for the strangeness $\mathrm{S=-3}$ and $\mathrm{-4}$ baryon–baryon
  interactions in chiral effective field theory}, Phys. Lett. B 684~(4) (2010)
  275 -- 280.
\newblock \href
  {https://doi.org/https://doi.org/10.1016/j.physletb.2010.01.031}
  {\path{doi:https://doi.org/10.1016/j.physletb.2010.01.031}}.
\newline\urlprefix\url{http://www.sciencedirect.com/science/article/pii/S0370269310000730}

\bibitem{Haidenbauer2013NPA915.0375}
J.~Haidenbauer, S.~Petschauer, N.~Kaiser, U.-G. Meißner, A.~Nogga, W.~Weise,
  \href{https://www.sciencedirect.com/science/article/pii/S0375947413006167}{Hyperon–nucleon
  interaction at next-to-leading order in chiral effective field theory}, Nucl.
  Phys. A 915 (2013) 24--58.
\newblock \href
  {https://doi.org/https://doi.org/10.1016/j.nuclphysa.2013.06.008}
  {\path{doi:https://doi.org/10.1016/j.nuclphysa.2013.06.008}}.
\newline\urlprefix\url{https://www.sciencedirect.com/science/article/pii/S0375947413006167}

\bibitem{Haidenbauer2016NPA954.273}
J.~Haidenbauer, U.-G. Meißner, S.~Petschauer,
  \href{https://www.sciencedirect.com/science/article/pii/S0375947416000075}{Strangeness
  $\mathrm{S}=\ensuremath{-}2$ baryon–baryon interaction at next-to-leading
  order in chiral effective field theory}, Nuclear Physics A 954 (2016)
  273--293.
\newblock \href
  {https://doi.org/https://doi.org/10.1016/j.nuclphysa.2016.01.006}
  {\path{doi:https://doi.org/10.1016/j.nuclphysa.2016.01.006}}.
\newline\urlprefix\url{https://www.sciencedirect.com/science/article/pii/S0375947416000075}

\bibitem{KaiWen2018CPC42.014105}
K.~W. Li, X.~L. Ren, L.~S. Geng, B.~W. Long,
  \href{http://hepnp.ihep.ac.cn//article/id/1e62084d-d7ac-42c0-b104-a107090e73c4}{Leading
  order relativistic hyperon-nucleon interactions in chiral effective field
  theory}, Chin. Phys. C 42 (2018) 014105.
\newblock \href {https://doi.org/10.1088/1674-1137/42/1/014105}
  {\path{doi:10.1088/1674-1137/42/1/014105}}.
\newline\urlprefix\url{http://hepnp.ihep.ac.cn//article/id/1e62084d-d7ac-42c0-b104-a107090e73c4}

\bibitem{KaiWen2018PRC98.065203}
K.~W. Li, T.~Hyodo, L.~S. Geng,
  \href{https://link.aps.org/doi/10.1103/PhysRevC.98.065203}{Strangeness
  $\mathrm{S}=\ensuremath{-}2$ baryon-baryon interactions in relativistic
  chiral effective field theory}, Phys. Rev. C 98 (2018) 065203.
\newblock \href {https://doi.org/10.1103/PhysRevC.98.065203}
  {\path{doi:10.1103/PhysRevC.98.065203}}.
\newline\urlprefix\url{https://link.aps.org/doi/10.1103/PhysRevC.98.065203}

\bibitem{Liu2021PRC103.025201}
Z.~W. Liu, J.~Song, K.~W. Li, L.~S. Geng,
  \href{https://link.aps.org/doi/10.1103/PhysRevC.103.025201}{Strangeness
  $\mathrm{S}=\ensuremath{-}3$ and $\mathrm{S}=\ensuremath{-}4$ baryon-baryon
  interactions in relativistic chiral effective field theory}, Phys. Rev. C 103
  (2021) 025201.
\newblock \href {https://doi.org/10.1103/PhysRevC.103.025201}
  {\path{doi:10.1103/PhysRevC.103.025201}}.
\newline\urlprefix\url{https://link.aps.org/doi/10.1103/PhysRevC.103.025201}

\bibitem{Goldhaber1960PR120.300}
G.~Goldhaber, S.~Goldhaber, W.~Lee, A.~Pais,
  \href{https://link.aps.org/doi/10.1103/PhysRev.120.300}{Influence of
  bose-einstein statistics on the antiproton-proton annihilation process},
  Phys. Rev. 120 (1960) 300--312.
\newblock \href {https://doi.org/10.1103/PhysRev.120.300}
  {\path{doi:10.1103/PhysRev.120.300}}.
\newline\urlprefix\url{https://link.aps.org/doi/10.1103/PhysRev.120.300}

\bibitem{Fung1978PRL41.1592}
S.~Y. Fung, W.~Gorn, G.~P. Kiernan, J.~J. Lu, Y.~T. Oh, R.~T. Poe,
  \href{https://link.aps.org/doi/10.1103/PhysRevLett.41.1592}{Observation of
  pion interferometry in relativistic nuclear collisions}, Phys. Rev. Lett. 41
  (1978) 1592--1594.
\newblock \href {https://doi.org/10.1103/PhysRevLett.41.1592}
  {\path{doi:10.1103/PhysRevLett.41.1592}}.
\newline\urlprefix\url{https://link.aps.org/doi/10.1103/PhysRevLett.41.1592}

\bibitem{Zajc1984PRC29.2173}
W.~A. Zajc, J.~A. Bistirlich, R.~R. Bossingham, H.~R. Bowman, C.~W. Clawson,
  K.~M. Crowe, K.~A. Frankel, J.~G. Ingersoll, J.~M. Kurck, C.~J. Martoff,
  D.~L. Murphy, J.~O. Rasmussen, J.~P. Sullivan, E.~Yoo, O.~Hashimoto,
  M.~Koike, W.~J. McDonald, J.~P. Miller, P.~Tru\"ol,
  \href{https://link.aps.org/doi/10.1103/PhysRevC.29.2173}{Two-pion
  correlations in heavy ion collisions}, Phys. Rev. C 29 (1984) 2173--2187.
\newblock \href {https://doi.org/10.1103/PhysRevC.29.2173}
  {\path{doi:10.1103/PhysRevC.29.2173}}.
\newline\urlprefix\url{https://link.aps.org/doi/10.1103/PhysRevC.29.2173}

\bibitem{Bamberger1988PLB203.320}
A.~Bamberger, et~al.,
  \href{https://www.sciencedirect.com/science/article/pii/0370269388905618}{Probing
  the space-time geometry of ultra-relativistic heavy-ion collisions}, Phys.
  Lett. B 203~(3) (1988) 320--326.
\newblock \href {https://doi.org/10.1016/0370-2693(88)90561-8}
  {\path{doi:10.1016/0370-2693(88)90561-8}}.
\newline\urlprefix\url{https://www.sciencedirect.com/science/article/pii/0370269388905618}

\bibitem{Podgoretsky1989FECAY20.628}
M.~I. Podgoretsky, Interference correlations of identical pions: Theory. (in
  russian), Fiz. Elem. Chast. Atom. Yadra 20 (1989) 628--668.

\bibitem{Abbott1992PRL69.1030}
T.~Abbott, et~al.,
  \href{https://link.aps.org/doi/10.1103/PhysRevLett.69.1030}{Bose-einstein
  correlations in $\mathrm{Si}$+$\mathrm{Al}$ and $\mathrm{Si}$+$\mathrm{Au}$
  collisions at 14.6$\mathrm{A GeV/c}$}, Phys. Rev. Lett. 69 (1992) 1030--1033.
\newblock \href {https://doi.org/10.1103/PhysRevLett.69.1030}
  {\path{doi:10.1103/PhysRevLett.69.1030}}.
\newline\urlprefix\url{https://link.aps.org/doi/10.1103/PhysRevLett.69.1030}

\bibitem{Barrette1994PLB333.33}
J.~Barrette, et~al.,
  \href{https://www.sciencedirect.com/science/article/pii/0370269394910049}{Evidence
  for expansion of a hot fireball from two-pion correlations for $\mathrm{Si}$
  + $\mathrm{Pb}$ collisions at $\mathrm{AGS}$ energy}, Phys. Lett. B 333~(1)
  (1994) 33--38.
\newblock \href {https://doi.org/10.1016/0370-2693(94)91004-9}
  {\path{doi:10.1016/0370-2693(94)91004-9}}.
\newline\urlprefix\url{https://www.sciencedirect.com/science/article/pii/0370269394910049}

\bibitem{Wiedemann1999PR319.145}
U.~A. Wiedemann, U.~Heinz,
  \href{https://www.sciencedirect.com/science/article/pii/S0370157399000320}{Particle
  interferometry for relativistic heavy-ion collisions}, Phys. Rep. 319~(4)
  (1999) 145--230.
\newblock \href {https://doi.org/10.1016/S0370-1573(99)00032-0}
  {\path{doi:10.1016/S0370-1573(99)00032-0}}.
\newline\urlprefix\url{https://www.sciencedirect.com/science/article/pii/S0370157399000320}

\bibitem{CMS2010PRL105.032001}
V.~Khachatryan, et~al.,
  \href{https://link.aps.org/doi/10.1103/PhysRevLett.105.032001}{First
  measurement of bose-einstein correlations in proton-proton collisions at
  $\sqrt{s}=0.9$ and 2.36 $\mathrm{TeV}$ at the $\mathrm{LHC}$}, Phys. Rev.
  Lett. 105 (2010) 032001.
\newblock \href {https://doi.org/10.1103/PhysRevLett.105.032001}
  {\path{doi:10.1103/PhysRevLett.105.032001}}.
\newline\urlprefix\url{https://link.aps.org/doi/10.1103/PhysRevLett.105.032001}

\bibitem{Koonin1977PLB70.43}
S.~E. Koonin,
  \href{https://www.sciencedirect.com/science/article/pii/0370269377903409}{Proton
  pictures of high-energy nuclear collisions}, Phys. Lett. B 70~(1) (1977)
  43--47.
\newblock \href {https://doi.org/10.1016/0370-2693(77)90340-9}
  {\path{doi:10.1016/0370-2693(77)90340-9}}.
\newline\urlprefix\url{https://www.sciencedirect.com/science/article/pii/0370269377903409}

\bibitem{Gyulassy1979PRC20.2267}
M.~Gyulassy, S.~K. Kauffmann, L.~W. Wilson,
  \href{https://link.aps.org/doi/10.1103/PhysRevC.20.2267}{Pion interferometry
  of nuclear collisions. $\mathrm{I.}$ theory}, Phys. Rev. C 20 (1979)
  2267--2292.
\newblock \href {https://doi.org/10.1103/PhysRevC.20.2267}
  {\path{doi:10.1103/PhysRevC.20.2267}}.
\newline\urlprefix\url{https://link.aps.org/doi/10.1103/PhysRevC.20.2267}

\bibitem{Lednicky1981YF35.1316}
R.~Lednicky, V.~L. Lyuboshits, Final state interaction effect on pairing
  correlations between particles with small relative momenta, Yad. Fiz. 35
  (1981) 1316--1330.

\bibitem{Pratt1990PRC42.2646}
S.~Pratt, T.~Cs\"org\ifmmode~\mbox{\H{o}}\else \H{o}\fi{}, J.~Zim\'anyi,
  \href{https://link.aps.org/doi/10.1103/PhysRevC.42.2646}{Detailed predictions
  for two-pion correlations in ultrarelativistic heavy-ion collisions}, Phys.
  Rev. C 42 (1990) 2646--2652.
\newblock \href {https://doi.org/10.1103/PhysRevC.42.2646}
  {\path{doi:10.1103/PhysRevC.42.2646}}.
\newline\urlprefix\url{https://link.aps.org/doi/10.1103/PhysRevC.42.2646}

\bibitem{Bauer1992ARNPS42.77}
W.~Bauer, C.~Gelbke, S.~Pratt,
  \href{https://doi.org/10.1146/annurev.ns.42.120192.000453}{Hadronic
  interferometry in heavy-ion collisions}, Annu. Rev. Nucl. Part. Sci. 42~(1)
  (1992) 77--98.
\newblock \href {https://doi.org/10.1146/annurev.ns.42.120192.000453}
  {\path{doi:10.1146/annurev.ns.42.120192.000453}}.
\newline\urlprefix\url{https://doi.org/10.1146/annurev.ns.42.120192.000453}

\bibitem{ALICE2017PRC96.064613}
S.~Acharya, et~al.,
  \href{https://link.aps.org/doi/10.1103/PhysRevC.96.064613}{Kaon femtoscopy in
  $\mathrm{Pb}$–$\mathrm{Pb}$ collisions at $\sqrt{{s}_{\mathit{NN}}}=2.76$
  $\mathrm{TeV}$}, Phys. Rev. C 96 (2017) 064613.
\newblock \href {https://doi.org/10.1103/PhysRevC.96.064613}
  {\path{doi:10.1103/PhysRevC.96.064613}}.
\newline\urlprefix\url{https://link.aps.org/doi/10.1103/PhysRevC.96.064613}

\bibitem{ALICE2017PLB774.64}
S.~Acharya, et~al.,
  \href{https://www.sciencedirect.com/science/article/pii/S0370269317307074}{Measuring
  $\mathrm{K_S^0K^{\pm}}$ interactions using $\mathrm{Pb}$–$\mathrm{Pb}$
  collisions at $\sqrt{{s}_{NN}}=2.76$ $\mathrm{TeV}$}, Phys. Lett. B 774
  (2017) 64--77.
\newblock \href {https://doi.org/10.1016/j.physletb.2017.09.009}
  {\path{doi:10.1016/j.physletb.2017.09.009}}.
\newline\urlprefix\url{https://www.sciencedirect.com/science/article/pii/S0370269317307074}

\bibitem{ALICE2019PLB790.22}
S.~Acharya, et~al.,
  \href{https://www.sciencedirect.com/science/article/pii/S0370269318309602}{Measuring
  $\mathrm{K_S^0K^{\pm}}$ interactions using $\mathrm{p}$–$\mathrm{p}$
  collisions at $\sqrt{{s}}=7$ $\mathrm{TeV}$}, Phys. Lett. B 790 (2019)
  22--34.
\newblock \href {https://doi.org/10.1016/j.physletb.2018.12.033}
  {\path{doi:10.1016/j.physletb.2018.12.033}}.
\newline\urlprefix\url{https://www.sciencedirect.com/science/article/pii/S0370269318309602}

\bibitem{ALICE2021arXiv2111.06611}
S.~Acharya, et~al.,
  \href{https://arxiv.org/abs/2111.06611}{$\mathrm{K_S^0K_S^0}$ and
  $\mathrm{K_S^0K^{\pm}}$ femtoscopy in $\mathrm{p}$-$\mathrm{p}$ collisions at
  $\sqrt{s}=5.02$ and $7$$\phantom{\rule{0.16em}{0ex}}\mathrm{TeV}$}, arXiv
  2111 (2021) 06611.
\newline\urlprefix\url{https://arxiv.org/abs/2111.06611}

\bibitem{ALICE2021PLB813.136030}
S.~Acharya, et~al.,
  \href{https://www.sciencedirect.com/science/article/pii/S0370269320308339}{Pion–kaon
  femtoscopy and the lifetime of the hadronic phase in
  $\mathrm{Pb}$−$\mathrm{Pb}$ collisions at $\sqrt{{s}_{NN}}=2.76$
  $\mathrm{TeV}$}, Phys. Lett. B 813 (2021) 136030.
\newblock \href {https://doi.org/10.1016/j.physletb.2020.136030}
  {\path{doi:10.1016/j.physletb.2020.136030}}.
\newline\urlprefix\url{https://www.sciencedirect.com/science/article/pii/S0370269320308339}

\bibitem{ALICE2020PRL124.092301}
S.~Acharya, et~al.,
  \href{https://link.aps.org/doi/10.1103/PhysRevLett.124.092301}{Scattering
  studies with low-energy kaon-proton femtoscopy in proton-proton collisions at
  the $\mathrm{LHC}$}, Phys. Rev. Lett. 124 (2020) 092301.
\newblock \href {https://doi.org/10.1103/PhysRevLett.124.092301}
  {\path{doi:10.1103/PhysRevLett.124.092301}}.
\newline\urlprefix\url{https://link.aps.org/doi/10.1103/PhysRevLett.124.092301}

\bibitem{ALICE2021PRC103.055201}
S.~Acharya, et~al.,
  \href{https://link.aps.org/doi/10.1103/PhysRevC.103.055201}{$\mathrm{\ensuremath{\Lambda}}\mathrm{K}$
  femtoscopy in $\mathrm{Pb}$-$\mathrm{Pb}$ collisions at
  $\sqrt{{s}_{NN}}=2.76$ $\mathrm{TeV}$}, Phys. Rev. C 103 (2021) 055201.
\newblock \href {https://doi.org/10.1103/PhysRevC.103.055201}
  {\path{doi:10.1103/PhysRevC.103.055201}}.
\newline\urlprefix\url{https://link.aps.org/doi/10.1103/PhysRevC.103.055201}

\bibitem{ALICE2021PRL127.172301}
S.~Acharya, et~al.,
  \href{https://link.aps.org/doi/10.1103/PhysRevLett.127.172301}{Experimental
  evidence for an attractive
  $\mathrm{\ensuremath{p}}$-$\mathrm{\ensuremath{\phi}}$ interaction}, Phys.
  Rev. Lett. 127 (2021) 172301.
\newblock \href {https://doi.org/10.1103/PhysRevLett.127.172301}
  {\path{doi:10.1103/PhysRevLett.127.172301}}.
\newline\urlprefix\url{https://link.aps.org/doi/10.1103/PhysRevLett.127.172301}

\bibitem{STAR2015PRL114.022301}
L.~Adamczyk, et~al.,
  \href{https://link.aps.org/doi/10.1103/PhysRevLett.114.022301}{$\mathrm{\ensuremath{\Lambda}}\mathrm{\ensuremath{\Lambda}}$
  correlation function in $\mathrm{Au}$+$\mathrm{Au}$ collisions at
  $\sqrt{{s}_{NN}}=200\text{ }\mathrm{GeV}$}, Phys. Rev. Lett. 114 (2015)
  022301.
\newblock \href {https://doi.org/10.1103/PhysRevLett.114.022301}
  {\path{doi:10.1103/PhysRevLett.114.022301}}.
\newline\urlprefix\url{https://link.aps.org/doi/10.1103/PhysRevLett.114.022301}

\bibitem{STAR2015Nature527.345}
L.~Adamczyk, et~al., \href{https://doi.org/10.1038/nature15724}{Measurement of
  interaction between antiprotons}, Nature 527 (2015) 345--348.
\newblock \href {https://doi.org/10.1038/nature15724}
  {\path{doi:10.1038/nature15724}}.
\newline\urlprefix\url{https://doi.org/10.1038/nature15724}

\bibitem{HADES2016PRC94.025201}
J.~Adamczewski-Musch, et~al.,
  \href{https://link.aps.org/doi/10.1103/PhysRevC.94.025201}{$\mathrm{\ensuremath{\Lambda}}\mathrm{p}$
  interaction studied via femtoscopy in
  $\mathrm{p}\phantom{\rule{0.28em}{0ex}}$+$\phantom{\rule{0.28em}{0ex}}\mathrm{Nb}$
  reactions at $\sqrt{{s}_{\mathit{NN}}}=3.18$ $\mathrm{GeV}$}, Phys. Rev. C 94
  (2016) 025201.
\newblock \href {https://doi.org/10.1103/PhysRevC.94.025201}
  {\path{doi:10.1103/PhysRevC.94.025201}}.
\newline\urlprefix\url{https://link.aps.org/doi/10.1103/PhysRevC.94.025201}

\bibitem{ALICE2019PRC99.024001}
S.~Acharya, et~al.,
  \href{https://link.aps.org/doi/10.1103/PhysRevC.99.024001}{$\mathrm{\ensuremath{p}}$-$\mathrm{\ensuremath{p}}$,
  $\mathrm{\ensuremath{p}}$-$\mathrm{\ensuremath{\Lambda}}$, and
  $\mathrm{\ensuremath{\Lambda}}$-$\mathrm{\ensuremath{\Lambda}}$ correlations
  studied via femtoscopy in $\mathrm{pp}$ reactions at
  $\sqrt{s}=7\phantom{\rule{0.16em}{0ex}}\mathrm{TeV}$}, Phys. Rev. C 99 (2019)
  024001.
\newblock \href {https://doi.org/10.1103/PhysRevC.99.024001}
  {\path{doi:10.1103/PhysRevC.99.024001}}.
\newline\urlprefix\url{https://link.aps.org/doi/10.1103/PhysRevC.99.024001}

\bibitem{ALICE2019PLB797.134822}
S.~Acharya, et~al.,
  \href{https://www.sciencedirect.com/science/article/pii/S0370269319305362}{Study
  of the $\mathrm{\ensuremath{\Lambda}}$-$\mathrm{\ensuremath{\Lambda}}$
  interaction with femtoscopy correlations in $\mathrm{pp}$ and
  $\mathrm{p}$–$\mathrm{Pb}$ collisions at the $\mathrm{LHC}$}, Phys. Lett. B
  797 (2019) 134822.
\newblock \href {https://doi.org/10.1016/j.physletb.2019.134822}
  {\path{doi:10.1016/j.physletb.2019.134822}}.
\newline\urlprefix\url{https://www.sciencedirect.com/science/article/pii/S0370269319305362}

\bibitem{STAR2019PLB790.490}
J.~Adam, et~al.,
  \href{https://www.sciencedirect.com/science/article/pii/S0370269319300802}{The
  proton–$\mathrm{\ensuremath{\Omega}}$ correlation function in
  $\mathrm{Au}$+$\mathrm{Au}$ collisions at $\sqrt{{s}_{NN}}=200\text{
  }\mathrm{GeV}$}, Phys. Lett. B 790 (2019) 490--497.
\newblock \href {https://doi.org/10.1016/j.physletb.2019.01.055}
  {\path{doi:10.1016/j.physletb.2019.01.055}}.
\newline\urlprefix\url{https://www.sciencedirect.com/science/article/pii/S0370269319300802}

\bibitem{ALICE2019PRL123.112002}
S.~Acharya, et~al.,
  \href{https://link.aps.org/doi/10.1103/PhysRevLett.123.112002}{First
  observation of an attractive interaction between a proton and a cascade
  baryon}, Phys. Rev. Lett. 123 (2019) 112002.
\newblock \href {https://doi.org/10.1103/PhysRevLett.123.112002}
  {\path{doi:10.1103/PhysRevLett.123.112002}}.
\newline\urlprefix\url{https://link.aps.org/doi/10.1103/PhysRevLett.123.112002}

\bibitem{ALICE2020PLB811.135849}
S.~Acharya, et~al.,
  \href{https://www.sciencedirect.com/science/article/pii/S0370269320306523}{Search
  for a common baryon source in high-multiplicity $\mathrm{p}$-$\mathrm{p}$
  collisions at the $\mathrm{LHC}$}, Phys. Lett. B 811 (2020) 135849.
\newblock \href {https://doi.org/10.1016/j.physletb.2020.135849}
  {\path{doi:10.1016/j.physletb.2020.135849}}.
\newline\urlprefix\url{https://www.sciencedirect.com/science/article/pii/S0370269320306523}

\bibitem{ALICE2020PLB805.135419}
S.~Acharya, et~al.,
  \href{https://www.sciencedirect.com/science/article/pii/S0370269320302239}{Investigation
  of the $\mathrm{\ensuremath{p}}$-$\mathrm{\ensuremath{\Sigma^0}}$ interaction
  via femtoscopy in $\mathrm{pp}$ collisions}, Phys. Lett. B 805 (2020) 135419.
\newblock \href {https://doi.org/10.1016/j.physletb.2020.135419}
  {\path{doi:10.1016/j.physletb.2020.135419}}.
\newline\urlprefix\url{https://www.sciencedirect.com/science/article/pii/S0370269320302239}

\bibitem{ALICE2020PLB802.135223}
S.~Acharya, et~al.,
  \href{https://www.sciencedirect.com/science/article/pii/S0370269320300277}{Measurement
  of strange baryon–antibaryon interactions with femtoscopic correlations},
  Phys. Lett. B 802 (2020) 135223.
\newblock \href {https://doi.org/10.1016/j.physletb.2020.135223}
  {\path{doi:10.1016/j.physletb.2020.135223}}.
\newline\urlprefix\url{https://www.sciencedirect.com/science/article/pii/S0370269320300277}

\bibitem{ALICE2020Nature588.232}
S.~Acharya, et~al., \href{https://doi.org/10.1038/s41586-020-3001-6}{Unveiling
  the strong interaction among hadrons at the $\mathrm{LHC}$}, Nature 588
  (2020) 232--238.
\newblock \href {https://doi.org/10.1038/s41586-020-3001-6}
  {\path{doi:10.1038/s41586-020-3001-6}}.
\newline\urlprefix\url{https://doi.org/10.1038/s41586-020-3001-6}

\bibitem{ALICE2021ARNPS71.377}
L.~Fabbietti, V.~M. Sarti, O.~V. Doce,
  \href{https://doi.org/10.1146/annurev-nucl-102419-034438}{Study of the strong
  interaction among hadrons with correlations at the $\mathrm{LHC}$}, Annu.
  Rev. Nucl. Part. Sci. 71 (2021) 377--402.
\newblock \href {https://doi.org/10.1146/annurev-nucl-102419-034438}
  {\path{doi:10.1146/annurev-nucl-102419-034438}}.
\newline\urlprefix\url{https://doi.org/10.1146/annurev-nucl-102419-034438}

\bibitem{ALICE2021arXiv2104.04427}
S.~Acharya, et~al., \href{https://arxiv.org/abs/2104.04427}{Exploring the
  $\mathrm{N}\mathrm{\ensuremath{\Lambda}}$-$\mathrm{N}\mathrm{\ensuremath{\Sigma}}$
  coupled system with high precision correlation techniques at the
  $\mathrm{LHC}$}, arXiv 2104 (2021) 04427.
\newline\urlprefix\url{https://arxiv.org/abs/2104.04427}

\bibitem{ALICE2021arXiv2105.05190}
S.~Acharya, et~al., \href{https://arxiv.org/abs/2105.05190}{Investigating the
  role of strangeness in baryon−antibaryon annihilation at the
  $\mathrm{LHC}$}, arXiv 2105 (2021) 05190.
\newline\urlprefix\url{https://arxiv.org/abs/2105.05190}

\bibitem{STAR2022EPJWC295.11015}
M.~Isshiki, \href{https://doi.org/10.1051/epjconf/202225911015}{Measurements of
  $\mathrm{p}$-$\mathrm{\ensuremath{\Xi}}$,
  $\mathrm{\ensuremath{\Lambda}}$-$\mathrm{\ensuremath{\Lambda}}$, and
  $\mathrm{\ensuremath{\Xi}}$-$\mathrm{\ensuremath{\Xi}}$ correlation in
  $\mathrm{Au}$+$\mathrm{Au}$ collisions at $\sqrt{{s}_{NN}}=200\text{
  }\mathrm{GeV}$ at $\mathrm{RHIC}$-$\mathrm{STAR}$}, EPJ Web Conf. 259 (2022)
  11015.
\newblock \href {https://doi.org/10.1051/epjconf/202225911015}
  {\path{doi:10.1051/epjconf/202225911015}}.
\newline\urlprefix\url{https://doi.org/10.1051/epjconf/202225911015}

\bibitem{Ohnishi2000NPA670.297}
A.~Ohnishi, Y.~Hirata, Y.~Nara, S.~Shinmura, Y.~Akaishi,
  \href{https://www.sciencedirect.com/science/article/pii/S0375947400001172}{Can
  we extract lambda-lambda interaction from two-particle momentum
  correlation?}, Nucl. Phys. A 670~(1) (2000) 297--300.
\newblock \href {https://doi.org/10.1016/S0375-9474(00)00117-2}
  {\path{doi:10.1016/S0375-9474(00)00117-2}}.
\newline\urlprefix\url{https://www.sciencedirect.com/science/article/pii/S0375947400001172}

\bibitem{Ohnishi2015PRC91.024916}
K.~Morita, T.~Furumoto, A.~Ohnishi,
  \href{https://link.aps.org/doi/10.1103/PhysRevC.91.024916}{$\ensuremath{\Lambda}\ensuremath{\Lambda}$
  interaction from relativistic heavy-ion collisions}, Phys. Rev. C 91 (2015)
  024916.
\newblock \href {https://doi.org/10.1103/PhysRevC.91.024916}
  {\path{doi:10.1103/PhysRevC.91.024916}}.
\newline\urlprefix\url{https://link.aps.org/doi/10.1103/PhysRevC.91.024916}

\bibitem{Ohnishi2016PRC94.031901}
K.~Morita, A.~Ohnishi, F.~Etminan, T.~Hatsuda,
  \href{https://link.aps.org/doi/10.1103/PhysRevC.94.031901}{Probing
  multistrange dibaryons with proton-omega correlations in high-energy heavy
  ion collisions}, Phys. Rev. C 94 (2016) 031901.
\newblock \href {https://doi.org/10.1103/PhysRevC.94.031901}
  {\path{doi:10.1103/PhysRevC.94.031901}}.
\newline\urlprefix\url{https://link.aps.org/doi/10.1103/PhysRevC.94.031901}

\bibitem{Ohnishi2016NPA954.294}
A.~Ohnishi, K.~Morita, K.~Miyahara, T.~Hyodo,
  \href{https://www.sciencedirect.com/science/article/pii/S0375947416301245}{Hadron–hadron
  correlation and interaction from heavy–ion collisions}, Nucl. Phys. A 954
  (2016) 294--307.
\newblock \href {https://doi.org/10.1016/j.nuclphysa.2016.05.010}
  {\path{doi:10.1016/j.nuclphysa.2016.05.010}}.
\newline\urlprefix\url{https://www.sciencedirect.com/science/article/pii/S0375947416301245}

\bibitem{Ohnishi2017PPNP95.279}
S.~Cho, T.~Hyodo, D.~Jido, C.~M. Ko, S.~H. Lee, S.~Maeda, K.~Miyahara,
  K.~Morita, M.~Nielsen, A.~Ohnishi, T.~Sekihara, T.~Song, S.~Yasui, K.~Yazaki,
  \href{https://www.sciencedirect.com/science/article/pii/S0146641017300182}{Exotic
  hadrons from heavy ion collisions}, Prog. Part. Nucl. Phys. 95 (2017)
  279--322.
\newblock \href {https://doi.org/10.1016/j.ppnp.2017.02.002}
  {\path{doi:10.1016/j.ppnp.2017.02.002}}.
\newline\urlprefix\url{https://www.sciencedirect.com/science/article/pii/S0146641017300182}

\bibitem{ALICE2018EPJC78.394}
D.~L. Mihaylov, V.~Mantovani~Sarti, O.~W. Arnold, L.~Fabbietti, B.~Hohlweger,
  A.~M. Mathis, \href{https://doi.org/10.1140/epjc/s10052-018-5859-0}{A
  femtoscopic correlation analysis tool using the $\mathrm{Schr\ddot odinger}$
  equation ($\mathrm{CATS}$)}, Eur. Phys. J. C 78 (2018) 394.
\newblock \href {https://doi.org/10.1140/epjc/s10052-018-5859-0}
  {\path{doi:10.1140/epjc/s10052-018-5859-0}}.
\newline\urlprefix\url{https://doi.org/10.1140/epjc/s10052-018-5859-0}

\bibitem{Ohnishi2020PRC101.015201}
K.~Morita, S.~Gongyo, T.~Hatsuda, T.~Hyodo, Y.~Kamiya, A.~Ohnishi,
  \href{https://link.aps.org/doi/10.1103/PhysRevC.101.015201}{Probing
  $\mathrm{\ensuremath{\Omega}}\mathrm{\ensuremath{\Omega}}$ and
  $\mathrm{p}\mathrm{\ensuremath{\Omega}}$ dibaryons with femtoscopic
  correlations in relativistic heavy-ion collisions}, Phys. Rev. C 101 (2020)
  015201.
\newblock \href {https://doi.org/10.1103/PhysRevC.101.015201}
  {\path{doi:10.1103/PhysRevC.101.015201}}.
\newline\urlprefix\url{https://link.aps.org/doi/10.1103/PhysRevC.101.015201}

\bibitem{Ohnishi2020PRL124.132501}
Y.~Kamiya, T.~Hyodo, K.~Morita, A.~Ohnishi, W.~Weise,
  \href{https://link.aps.org/doi/10.1103/PhysRevLett.124.132501}{$\mathrm{K^-}\mathrm{p}$
  correlation function from high-energy nuclear collisions and chiral
  $\mathrm{SU}$(3) dynamics}, Phys. Rev. Lett. 124 (2020) 132501.
\newblock \href {https://doi.org/10.1103/PhysRevLett.124.132501}
  {\path{doi:10.1103/PhysRevLett.124.132501}}.
\newline\urlprefix\url{https://link.aps.org/doi/10.1103/PhysRevLett.124.132501}

\bibitem{Ohnishi2021PRC105.014915}
Y.~Kamiya, K.~Sasaki, T.~Fukui, T.~Hyodo, K.~Morita, K.~Ogata, A.~Ohnishi,
  T.~Hatsuda,
  \href{https://link.aps.org/doi/10.1103/PhysRevC.105.014915}{Femtoscopic study
  of coupled-channels $\mathrm{N}\mathrm{\ensuremath{\Xi}}$ and
  $\mathrm{\ensuremath{\Lambda}}\mathrm{\ensuremath{\Lambda}}$ interactions},
  Phys. Rev. C 105 (2022) 014915.
\newblock \href {https://doi.org/10.1103/PhysRevC.105.014915}
  {\path{doi:10.1103/PhysRevC.105.014915}}.
\newline\urlprefix\url{https://link.aps.org/doi/10.1103/PhysRevC.105.014915}

\bibitem{Ohnishi2021FBS62.42}
A.~Ohnishi, Y.~Kamiya, K.~Sasaki, T.~Fukui, T.~Hatsuda, T.~Hyodo, K.~Morita,
  K.~Ogata, \href{https://doi.org/10.1007/s00601-021-01626-z}{Femtoscopic study
  of $\mathrm{N}\mathrm{\ensuremath{\Xi}}$ interaction and search for the
  $\mathrm{H}$ dibaryon state around the $\mathrm{N}\mathrm{\ensuremath{\Xi}}$
  threshold}, Few-Body Syst. 62 (2021) 42.
\newblock \href {https://doi.org/10.1007/s00601-021-01626-z}
  {\path{doi:10.1007/s00601-021-01626-z}}.
\newline\urlprefix\url{https://doi.org/10.1007/s00601-021-01626-z}

\bibitem{Ohnishi2021PRC103.065205}
K.~Ogata, T.~Fukui, Y.~Kamiya, A.~Ohnishi,
  \href{https://link.aps.org/doi/10.1103/PhysRevC.103.065205}{Effect of
  deuteron breakup on the deuteron-$\mathrm{\ensuremath{\Xi}}$ correlation
  function}, Phys. Rev. C 103 (2021) 065205.
\newblock \href {https://doi.org/10.1103/PhysRevC.103.065205}
  {\path{doi:10.1103/PhysRevC.103.065205}}.
\newline\urlprefix\url{https://link.aps.org/doi/10.1103/PhysRevC.103.065205}

\bibitem{Haidenbauer2019NPA981.1}
J.~Haidenbauer,
  \href{https://www.sciencedirect.com/science/article/pii/S0375947418303774}{Coupled-channel
  effects in hadron–hadron correlation functions}, Nucl. Phys. A 981 (2019)
  1--16.
\newblock \href {https://doi.org/10.1016/j.nuclphysa.2018.10.090}
  {\path{doi:10.1016/j.nuclphysa.2018.10.090}}.
\newline\urlprefix\url{https://www.sciencedirect.com/science/article/pii/S0375947418303774}

\bibitem{Haidenbauer2020EPJA56.184}
J.~Haidenbauer, G.~Krein, T.~C. Peixoto,
  \href{https://doi.org/10.1140/epja/s10050-020-00190-0}{Femtoscopic
  correlations and the $\mathrm{\ensuremath{\Lambda _c}}\mathrm{N}$
  interaction}, Eur. Phys. J. A 56 (2020) 184.
\newblock \href {https://doi.org/10.1140/epja/s10050-020-00190-0}
  {\path{doi:10.1140/epja/s10050-020-00190-0}}.
\newline\urlprefix\url{https://doi.org/10.1140/epja/s10050-020-00190-0}

\bibitem{Haidenbauer2022PLB829.137074}
J.~Haidenbauer, U.~G. Mei\ss{}ner,
  \href{https://www.sciencedirect.com/science/article/pii/S0370269322002088}{Exploring
  the $\mathrm{\ensuremath{\Sigma^+}}\mathrm{\ensuremath{p}}$ interaction by
  measurements of the correlation function}, Phys. Lett. B 829 (2022) 137074.
\newblock \href
  {https://doi.org/https://doi.org/10.1016/j.physletb.2022.137074}
  {\path{doi:https://doi.org/10.1016/j.physletb.2022.137074}}.
\newline\urlprefix\url{https://www.sciencedirect.com/science/article/pii/S0370269322002088}

\bibitem{Graczykowski2021PRC104.054909}
L.~K. Graczykowski, M.~A. Janik,
  \href{https://link.aps.org/doi/10.1103/PhysRevC.104.054909}{Unfolding the
  effects of final-state interactions and quantum statistics in two-particle
  angular correlations}, Phys. Rev. C 104 (2021) 054909.
\newblock \href {https://doi.org/10.1103/PhysRevC.104.054909}
  {\path{doi:10.1103/PhysRevC.104.054909}}.
\newline\urlprefix\url{https://link.aps.org/doi/10.1103/PhysRevC.104.054909}

\bibitem{Silverio2021arXiv2110.15455}
I.~M. Silv\'erio, S.~S. Padula, G.~a.~I. Krein,
  \href{https://arxiv.org/abs/2110.15455}{Study of $\mathrm{D^0}$ meson
  interactions via femtoscopic correlations}, arXiv 2110 (2021) 15455.
\newline\urlprefix\url{https://arxiv.org/abs/2110.15455}

\bibitem{Mrowczynski2021PRC104.024909}
S.~Mr\'owczy\ifmmode~\acute{n}\else \'{n}\fi{}ski,
  P.~S\l{}o\ifmmode~\acute{n}\else \'{n}\fi{},
  \href{https://link.aps.org/doi/10.1103/PhysRevC.104.024909}{Deuteron-deuteron
  correlation function in nucleus-nucleus collisions}, Phys. Rev. C 104 (2021)
  024909.
\newblock \href {https://doi.org/10.1103/PhysRevC.104.024909}
  {\path{doi:10.1103/PhysRevC.104.024909}}.
\newline\urlprefix\url{https://link.aps.org/doi/10.1103/PhysRevC.104.024909}

\bibitem{NPLQCD2007NPA794.62}
S.~R. Beane, P.~F. Bedaque, T.~C. Luu, K.~Orginos, E.~Pallante, A.~Parreño,
  M.~J. Savage,
  \href{https://www.sciencedirect.com/science/article/pii/S0375947407006434}{Hyperon–nucleon
  scattering from fully-dynamical lattice $\mathrm{QCD}$}, Nucl. Phys. A
  794~(1) (2007) 62--72.
\newblock \href
  {https://doi.org/https://doi.org/10.1016/j.nuclphysa.2007.07.006}
  {\path{doi:https://doi.org/10.1016/j.nuclphysa.2007.07.006}}.
\newline\urlprefix\url{https://www.sciencedirect.com/science/article/pii/S0375947407006434}

\bibitem{Nemura2018EPJWebConf175.05030}
H.~Nemura, S.~Aoki, T.~Doi, S.~Gongyo, T.~Hatsuda, Y.~Ikeda, T.~Inoue,
  T.~Iritani, N.~Ishii, T.~Miyamoto, K.~Sasaki,
  \href{https://doi.org/10.1051/epjconf/201817505030}{Baryon interactions from
  lattice $\mathrm{QCD}$ with physical masses — strangeness $\mathrm{S = -1}$
  sector}, EPJ Web Conf. 175 (2018) 05030.
\newblock \href {https://doi.org/10.1051/epjconf/201817505030}
  {\path{doi:10.1051/epjconf/201817505030}}.
\newline\urlprefix\url{https://doi.org/10.1051/epjconf/201817505030}

\bibitem{Ishii2018EPJWebConf175.05013}
N.~Ishii, S.~Aoki, T.~Doi, S.~Gongyo, T.~Hatsuda, Y.~Ikeda, T.~Inoue,
  T.~Iritani, T.~Miyamoto, H.~Nemura, K.~Sasaki,
  \href{https://doi.org/10.1051/epjconf/201817505013}{Baryon interactions from
  lattice $\mathrm{QCD}$ with physical masses —$\mathrm{S = −3}$ sector:
  $\mathrm{\Xi\Sigma}$ and $\mathrm{\Xi\Lambda-\Xi\Sigma}$}, EPJ Web Conf. 175
  (2018) 05013.
\newblock \href {https://doi.org/10.1051/epjconf/201817505013}
  {\path{doi:10.1051/epjconf/201817505013}}.
\newline\urlprefix\url{https://doi.org/10.1051/epjconf/201817505013}

\bibitem{Doi2018EPJWebConf175.05009}
T.~Doi, T.~Iritani, S.~Aoki, S.~Gongyo, T.~Hatsuda, Y.~Ikeda, T.~Inoue,
  N.~Ishii, T.~Miyamoto, H.~Nemura, K.~Sasaki,
  \href{https://doi.org/10.1051/epjconf/201817505009}{Baryon interactions from
  lattice $\mathrm{QCD}$ with physical quark masses – nuclear forces and
  $\mathrm{\Xi\Xi}$ forces}, EPJ Web Conf. 175 (2018) 05009.
\newblock \href {https://doi.org/10.1051/epjconf/201817505009}
  {\path{doi:10.1051/epjconf/201817505009}}.
\newline\urlprefix\url{https://doi.org/10.1051/epjconf/201817505009}

\bibitem{Sasaki2020NPA998.121737}
K.~Sasaki, S.~Aoki, T.~Doi, S.~Gongyo, T.~Hatsuda, Y.~Ikeda, T.~Inoue,
  T.~Iritani, N.~Ishii, K.~Murano, T.~Miyamoto,
  \href{https://www.sciencedirect.com/science/article/pii/S0375947420300476}{$\mathrm{\Lambda\Lambda}$
  and $\mathrm{\Xi}\mathrm{N}$ interactions from lattice $\mathrm{QCD}$ near
  the physical point}, Nucl. Phys. A 998 (2020) 121737.
\newblock \href {https://doi.org/10.1016/j.nuclphysa.2020.121737}
  {\path{doi:10.1016/j.nuclphysa.2020.121737}}.
\newline\urlprefix\url{https://www.sciencedirect.com/science/article/pii/S0375947420300476}

\bibitem{NPLQCD2021PRD103.054508}
M.~Illa, S.~R. Beane, E.~Chang, Z.~Davoudi, W.~Detmold, D.~J. Murphy,
  K.~Orginos, A.~Parre\~no, M.~J. Savage, P.~E. Shanahan, M.~L. Wagman,
  F.~Winter,
  \href{https://link.aps.org/doi/10.1103/PhysRevD.103.054508}{Low-energy
  scattering and effective interactions of two baryons at
  ${m}_{\ensuremath{\pi}}\ensuremath{\sim}450\text{ }\text{ }\mathrm{MeV}$ from
  lattice quantum chromodynamics}, Phys. Rev. D 103 (2021) 054508.
\newblock \href {https://doi.org/10.1103/PhysRevD.103.054508}
  {\path{doi:10.1103/PhysRevD.103.054508}}.
\newline\urlprefix\url{https://link.aps.org/doi/10.1103/PhysRevD.103.054508}

\bibitem{Ren2018CPC42.014103}
X.~L. Ren, K.~W. Li, L.~S. Geng, B.~W. Long, P.~Ring, J.~Meng,
  \href{https://doi.org/10.1088\%2F1674-1137\%2F42\%2F1\%2F014103}{Leading
  order relativistic chiral nucleon-nucleon interaction}, Chin. Phys. C 42
  (2018) 014103.
\newblock \href {https://doi.org/10.1088/1674-1137/42/1/014103}
  {\path{doi:10.1088/1674-1137/42/1/014103}}.
\newline\urlprefix\url{https://doi.org/10.1088\%2F1674-1137\%2F42\%2F1\%2F014103}

\bibitem{Xiao2019PRC99.024004}
Y.~Xiao, L.~S. Geng, X.~L. Ren,
  \href{https://link.aps.org/doi/10.1103/PhysRevC.99.024004}{Covariant
  nucleon-nucleon contact lagrangian up to order $\mathcal{O}({q}^{4})$}, Phys.
  Rev. C 99 (2019) 024004.
\newblock \href {https://doi.org/10.1103/PhysRevC.99.024004}
  {\path{doi:10.1103/PhysRevC.99.024004}}.
\newline\urlprefix\url{https://link.aps.org/doi/10.1103/PhysRevC.99.024004}

\bibitem{Xiao2020PRC102.054001}
Y.~Xiao, C.~X. Wang, J.~X. Lu, L.~S. Geng,
  \href{https://link.aps.org/doi/10.1103/PhysRevC.102.054001}{Two-pion exchange
  contributions to the nucleon-nucleon interaction in covariant baryon chiral
  perturbation theory}, Phys. Rev. C 102 (2020) 054001.
\newblock \href {https://doi.org/10.1103/PhysRevC.102.054001}
  {\path{doi:10.1103/PhysRevC.102.054001}}.
\newline\urlprefix\url{https://link.aps.org/doi/10.1103/PhysRevC.102.054001}

\bibitem{Bai2020PLB809.135745}
Q.~Q. Bai, C.~X. Wang, Y.~Xiao, L.~S. Geng,
  \href{https://www.sciencedirect.com/science/article/pii/S0370269320305487}{Pion-mass
  dependence of the nucleon-nucleon interaction}, Phys. Lett. B 809 (2020)
  135745.
\newblock \href {https://doi.org/10.1016/j.physletb.2020.135745}
  {\path{doi:10.1016/j.physletb.2020.135745}}.
\newline\urlprefix\url{https://www.sciencedirect.com/science/article/pii/S0370269320305487}

\bibitem{Bai2021arXiv2105.06113}
Q.~Q. Bai, C.~X. Wang, Y.~Xiao, L.~S. Geng,
  \href{https://arxiv.org/abs/2105.06113}{Pion-mass dependence of the
  nucleon-nucleon interaction in the $\mathrm{^3S_1}$-$\mathrm{^3D_1}$ coupled
  channel}, arXiv 2105 (2021) 06113.
\newline\urlprefix\url{https://arxiv.org/abs/2105.06113}

\bibitem{Wang2022PRC105.014003}
C.~X. Wang, J.~X. Lu, Y.~Xiao, L.~S. Geng,
  \href{https://link.aps.org/doi/10.1103/PhysRevC.105.014003}{Nonperturbative
  two-pion exchange contributions to the nucleon-nucleon interaction in
  covariant baryon chiral perturbation theory}, Phys. Rev. C 105 (2022) 014003.
\newblock \href {https://doi.org/10.1103/PhysRevC.105.014003}
  {\path{doi:10.1103/PhysRevC.105.014003}}.
\newline\urlprefix\url{https://link.aps.org/doi/10.1103/PhysRevC.105.014003}

\bibitem{Lu2022PRL128.142002}
J.~X. Lu, C.~X. Wang, Y.~Xiao, L.~S. Geng, J.~Meng, P.~Ring,
  \href{https://link.aps.org/doi/10.1103/PhysRevLett.128.142002}{Accurate
  relativistic chiral nucleon-nucleon interaction up to next-to-next-to-leading
  order}, Phys. Rev. Lett. 128 (2022) 142002.
\newblock \href {https://doi.org/10.1103/PhysRevLett.128.142002}
  {\path{doi:10.1103/PhysRevLett.128.142002}}.
\newline\urlprefix\url{https://link.aps.org/doi/10.1103/PhysRevLett.128.142002}

\bibitem{Ren2021CPL38.062101}
X.~L. Ren, C.~X. Wang, K.~W. Li, L.~S. Geng, J.~Meng,
  \href{https://doi.org/10.1088/0256-307x/38/6/062101}{Relativistic chiral
  description of the $\mathrm{^1S_0}$ nucleon{\textendash}nucleon scattering},
  Chin. Phys. Lett. 38 (2021) 062101.
\newblock \href {https://doi.org/10.1088/0256-307x/38/6/062101}
  {\path{doi:10.1088/0256-307x/38/6/062101}}.
\newline\urlprefix\url{https://doi.org/10.1088/0256-307x/38/6/062101}

\bibitem{Wang2021CPC45.054101}
C.~X. Wang, L.~S. Geng, B.~W. Long,
  \href{https://doi.org/10.1088/1674-1137/abe368}{Renormalizability of leading
  order covariant chiral nucleon-nucleon interaction}, Chin. Phys. C 45 (2021)
  054101.
\newblock \href {https://doi.org/10.1088/1674-1137/abe368}
  {\path{doi:10.1088/1674-1137/abe368}}.
\newline\urlprefix\url{https://doi.org/10.1088/1674-1137/abe368}

\bibitem{KaiWen2016PRD94.014029}
K.~W. Li, X.~L. Ren, L.~S. Geng, B.~W. Long,
  \href{https://link.aps.org/doi/10.1103/PhysRevD.94.014029}{Strangeness
  $\mathrm{S}=\ensuremath{-}1$ hyperon-nucleon scattering in covariant chiral
  effective field theory}, Phys. Rev. D 94 (2016) 014029.
\newblock \href {https://doi.org/10.1103/PhysRevD.94.014029}
  {\path{doi:10.1103/PhysRevD.94.014029}}.
\newline\urlprefix\url{https://link.aps.org/doi/10.1103/PhysRevD.94.014029}

\bibitem{Song2018PRC97.065201}
J.~Song, K.~W. Li, L.~S. Geng,
  \href{https://link.aps.org/doi/10.1103/PhysRevC.97.065201}{Strangeness
  $\mathrm{S}=\ensuremath{-}1$ hyperon-nucleon interactions: Chiral effective
  field theory versus lattice $\mathrm{QCD}$}, Phys. Rev. C 97 (2018) 065201.
\newblock \href {https://doi.org/10.1103/PhysRevC.97.065201}
  {\path{doi:10.1103/PhysRevC.97.065201}}.
\newline\urlprefix\url{https://link.aps.org/doi/10.1103/PhysRevC.97.065201}

\bibitem{Song2022PRC105.035203}
J.~Song, Z.-W. Liu, K.-W. Li, L.-S. Geng,
  \href{https://link.aps.org/doi/10.1103/PhysRevC.105.035203}{Test of the
  hyperon-nucleon interaction within leading order covariant chiral effective
  field theory}, Phys. Rev. C 105 (2022) 035203.
\newblock \href {https://doi.org/10.1103/PhysRevC.105.035203}
  {\path{doi:10.1103/PhysRevC.105.035203}}.
\newline\urlprefix\url{https://link.aps.org/doi/10.1103/PhysRevC.105.035203}

\bibitem{Song2020PRC102.065208}
J.~Song, Y.~Xiao, Z.~W. Liu, C.~X. Wang, K.~W. Li, L.~S. Geng,
  \href{https://link.aps.org/doi/10.1103/PhysRevC.102.065208}{${\mathrm{\ensuremath{\Lambda}}}_{c}\mathrm{N}$
  interaction in leading-order covariant chiral effective field theory}, Phys.
  Rev. C 102 (2020) 065208.
\newblock \href {https://doi.org/10.1103/PhysRevC.102.065208}
  {\path{doi:10.1103/PhysRevC.102.065208}}.
\newline\urlprefix\url{https://link.aps.org/doi/10.1103/PhysRevC.102.065208}

\bibitem{Song2021arXiv2104.02380}
J.~Song, Y.~Xiao, Z.~W. Liu, K.~W. Li, L.~S. Geng,
  \href{https://arxiv.org/abs/2104.02380}{$\mathrm{^3S_1}$-$\mathrm{{}^3D_1}$
  coupled channel $\mathrm{\Lambda_c}\mathrm{N}$ interactions: chiral effective
  field theory vs. lattice $\mathrm{QCD}$}, arXiv 2104 (2021) 02380.
\newline\urlprefix\url{https://arxiv.org/abs/2104.02380}

\bibitem{Haidenbauer2015EPJA51.17}
J.~Haidenbauer, U.-G. Meißner, S.~Petschauer,
  \href{https://doi.org/10.1140/epja/i2015-15017-0}{Do
  $\mathrm{\ensuremath{\Xi}}\mathrm{\ensuremath{\Xi}}$ bound states exist?},
  Eur. Phys. J. A 51 (2015) 17.
\newblock \href {https://doi.org/10.1140/epja/i2015-15017-0}
  {\path{doi:10.1140/epja/i2015-15017-0}}.
\newline\urlprefix\url{https://doi.org/10.1140/epja/i2015-15017-0}

\bibitem{Kadyshevsky1968NPB6.125}
V.~G. Kadyshevsky,
  \href{https://www.sciencedirect.com/science/article/pii/0550321368902745}{Quasipotential
  type equation for the relativistic scattering amplitude}, Nucl. Phys. B 6~(2)
  (1968) 125--148.
\newblock \href {https://doi.org/10.1016/0550-3213(68)90274-5}
  {\path{doi:10.1016/0550-3213(68)90274-5}}.
\newline\urlprefix\url{https://www.sciencedirect.com/science/article/pii/0550321368902745}

\bibitem{Sasaki2018EPJWebConf175.05010}
K.~Sasaki, S.~Aoki, T.~Doi, S.~Gongyo, T.~Hatsuda, Y.~Ikeda, T.~Inoue,
  T.~Iritani, N.~Ishii, T.~Miyamoto,
  \href{https://doi.org/10.1051/epjconf/201817505010}{Lattice $\mathrm{QCD}$
  studies on baryon interactions in the strangeness -2 sector with physical
  quark masses}, EPJ Web Conf. 175 (2018) 05010.
\newblock \href {https://doi.org/10.1051/epjconf/201817505010}
  {\path{doi:10.1051/epjconf/201817505010}}.
\newline\urlprefix\url{https://doi.org/10.1051/epjconf/201817505010}

\bibitem{Aoki2011PJASB87.509}
S.~Aoki, N.~Ishii, T.~Doi, T.~Hatsuda, Y.~Ikeda, T.~Inoue, K.~Murano,
  H.~Nemura, K.~Sasaki, \href{https://doi.org/10.2183/pjab.87.509}{Extraction
  of hadron interactions above inelastic threshold in lattice $\mathrm{QCD}$},
  Proc. Jpn. Acad. Ser. B 87~(8) (2011) 509--517.
\newblock \href {https://doi.org/10.2183/pjab.87.509}
  {\path{doi:10.2183/pjab.87.509}}.
\newline\urlprefix\url{https://doi.org/10.2183/pjab.87.509}

\bibitem{Aoki2013PRD87.034512}
S.~Aoki, B.~Charron, T.~Doi, T.~Hatsuda, T.~Inoue, N.~Ishii,
  \href{https://link.aps.org/doi/10.1103/PhysRevD.87.034512}{Construction of
  energy-independent potentials above inelastic thresholds in quantum field
  theories}, Phys. Rev. D 87 (2013) 034512.
\newblock \href {https://doi.org/10.1103/PhysRevD.87.034512}
  {\path{doi:10.1103/PhysRevD.87.034512}}.
\newline\urlprefix\url{https://link.aps.org/doi/10.1103/PhysRevD.87.034512}

\bibitem{Vincent1974PRC10.391}
C.~M. Vincent, S.~C. Phatak,
  \href{https://link.aps.org/doi/10.1103/PhysRevC.10.391}{Accurate
  momentum-space method for scattering by nuclear and $\mathrm{Coulomb}$
  potentials}, Phys. Rev. C 10 (1974) 391--394.
\newblock \href {https://doi.org/10.1103/PhysRevC.10.391}
  {\path{doi:10.1103/PhysRevC.10.391}}.
\newline\urlprefix\url{https://link.aps.org/doi/10.1103/PhysRevC.10.391}

\bibitem{Holzenkamp1989NPA500.485}
B.~Holzenkamp, K.~Holinde, J.~Speth,
  \href{https://www.sciencedirect.com/science/article/pii/0375947489902236}{A
  meson exchange model for the hyperon-nucleon interaction}, Nucl. Phys. A 500
  (1989) 485--528.
\newblock \href {https://doi.org/10.1016/0375-9474(89)90223-6}
  {\path{doi:10.1016/0375-9474(89)90223-6}}.
\newline\urlprefix\url{https://www.sciencedirect.com/science/article/pii/0375947489902236}

\bibitem{Bazavov2014PRD90.094503}
A.~Bazavov, et~al.,
  \href{https://link.aps.org/doi/10.1103/PhysRevD.90.094503}{Equation of state
  in ($2+1$)-flavor $\mathrm{QCD}$}, Phys. Rev. D 90 (2014) 094503.
\newblock \href {https://doi.org/10.1103/PhysRevD.90.094503}
  {\path{doi:10.1103/PhysRevD.90.094503}}.
\newline\urlprefix\url{https://link.aps.org/doi/10.1103/PhysRevD.90.094503}

\bibitem{Borsanyi2014PLB730.99}
S.~Borsányi, Z.~Fodor, C.~Hoelbling, S.~D. Katz, S.~Krieg, K.~K. Szabó,
  \href{https://www.sciencedirect.com/science/article/pii/S0370269314000197}{Full
  result for the $\mathrm{QCD}$ equation of state with 2+1 flavors}, Phys.
  Lett. B 730 (2014) 99--104.
\newblock \href {https://doi.org/10.1016/j.physletb.2014.01.007}
  {\path{doi:10.1016/j.physletb.2014.01.007}}.
\newline\urlprefix\url{https://www.sciencedirect.com/science/article/pii/S0370269314000197}

\bibitem{Oliver2017}
A.~Oliver~Werner, Study of the hyperon-nucleon interaction via femtoscopy in
  elementary systems with hades and alice, Dissertation, Technische
  Universität München, München (2017).

\bibitem{ALICE-PUBLIC-2020-005}
A.~Collaboration, \href{https://cds.cern.ch/record/2724925}{Future high-energy
  pp programme with $\mathrm{ALICE}$}, ALICE-PUBLIC-2020-005 (Jul 2020).
\newline\urlprefix\url{https://cds.cern.ch/record/2724925}

\bibitem{Miller2013CJP51.466}
G.~Miller, Detecting strangeness -4 dibaryon states, Chin. J. Phys. 51 (2013)
  466.
\newblock \href {https://doi.org/10.6122/CJP.51.466}
  {\path{doi:10.6122/CJP.51.466}}.

\bibitem{Ahn2011JPARC}
J.~K. Ahn, R.~Honda, Y.~Ichikawa, K.~Imai, R.~Kiuchi, H.~Sako,
  et~al.\url{http://j-parc.jp/researcher/Hadron/en/pac_1107/pdf/KEK_J-PARC-PAC2011-03.pdf}
  (2011).

\end{thebibliography}
\end{document}